\documentclass[sn-mathphys-num]{sn-jnl}

\usepackage{graphicx}%
\usepackage{multirow}%
\usepackage{amsmath,amssymb,amsfonts}%
\usepackage{amsthm}%
\usepackage{mathrsfs}%
\usepackage[title]{appendix}%
\usepackage{xcolor}%
\usepackage{algpseudocode}%
\usepackage{listings}%
\usepackage{dsfont}

\usepackage{bbold}
\usepackage{placeins}
\usepackage{subcaption}

\usepackage{tikz}
\usetikzlibrary{quantikz}

\newcommand\scalemath[2]{\scalebox{#1}{\mbox{\ensuremath{\displaystyle #2}}}}
\DeclarePairedDelimiter\abs{\lvert}{\rvert}%

\begin{document}

\title[Explicit Quantum Circuit for Simulating the Advection-Diffusion-Reaction Dynamics]{Explicit Quantum Circuit for Simulating the Advection-Diffusion-Reaction Dynamics}


\author*[1]{\fnm{Claudio} \sur{Sanavio}}\email{claudio.sanavio@iit.it}
\equalcont{These authors contributed equally to this work.}

\author[1]{\fnm{Enea} \sur{Mauri}}\email{emauri.research@proton.me}
\equalcont{These authors contributed equally to this work.}

\author[1]{\fnm{Sauro} \sur{Succi}}\email{sauro.succi@gmail.com}

\affil*[1]{\orgdiv{Center for Life Nano-Neuroscience at la Sapienza}, \orgname{Fondazione Istituto Italiano di Tecnologia}, \orgaddress{\street{Viale Regina Elena 291}, \city{Roma}, \postcode{00161}, \country{Italy}}}


\abstract{We assess the convergence of the Carleman linearization of advection-diffusion-reaction (ADR) equations
with a logistic nonlinearity. It is shown that five Carleman iterates provide a satisfactory approximation
of the original ADR across a broad range of parameters and strength of nonlinearity.
To assess the feasibility of a quantum algorithm based on this linearization, we analyze the projection of the Carleman ADR matrix onto the tensor 
Pauli basis.
It is found that the Carleman ADR matrix requires an exponential number of Pauli gates 
as a function of the number of qubits.
This prevents the practical implementation of the Carleman approach to the quantum
simulation of ADR problems on current hardware.
We propose to address this limitation by resorting to block-encoding techniques for sparse matrix
employing oracles.
Such quantum ADR oracles are presented in {\it explicit} form and shown to turn the exponential complexity into a polynomial one.
However, due to the low probability of successfully implementing the nonunitary Carleman operator, further research is needed to implement the multi--timestep version of the present circuit.}

\keywords{advection-diffusion-reaction, carleman linearization, quantum computing, block encoding}

\maketitle

\section{Introduction\label{sec:I}}

Quantum computing, the field of exploiting properties of quantum states to process 
information, could potentially lead to a revolution in our computational capabilities, overcoming 
the limitations of classical computers for a selected range of applications. 
Various areas can benefit from these advances, ranging from the simulation of the dynamics of quantum systems~\cite{feynman_simulating_1982}, to solving a set of linear 
equations\cite{harrow_quantum_2009,childs_quantum_2017}. 
Generally, improvements are expected for problems affected by 
the  ``curse of dimensionality", where the demand for computer resources grows exponentially 
with the size of the problem. However, other important problems may benefit as well, an outstanding
one being computational fluid dynamics (CFD) \cite{agrawal_reynolds-number-dependence_2024,steijl_quantum_2019,bharadwaj_quantum_2020,succi_ensemble_2024,sanavio_three_2024,sanavio_quantum_2024}. 

CFD is notoriously difficult because the computational complexity to study turbulent flows grows as $\sim \text{Re}^3$ \cite{succi_lattice_2018}, $\text{Re}$ 
being the Reynolds number, where ordinary fluid phenomena, such as the flow around the car, involve Reynolds number of the order of ten millions. 

One of the main advantages of a quantum algorithm for fluid dynamics would be a logarithmic reduction of these 
resources, as the classical information can be encoded into the exponentially higher dimension of the 
Hilbert space of the quantum states. In fact, $N$ variables can be encoded by using just $n = \log_2(N)$ qubits. 

It is estimated that with about eighty logical qubits one could outdo present-day exascale supercomputers \cite{succi_quantum_2023}. This would allow to overcome the limit of current simulation at high Reynolds numbers, for $\text{Re} > 10^7$~\cite{agrawal_reynolds-number-dependence_2024,falcucci_extreme_2021}. 

For this reason and due to the importance of the problem in a variety of scientific and engineering problems, quantum 
algorithms for fluid dynamics are attracting a lot of interest\cite{steijl_quantum_2022, schalkers_importance_2024, liu_efficient_2021,li_potential_2023, liu_efficient_2023}. 
Despite these valiant attempts, an explicit and efficient 
implementation of a quantum circuit for fluid simulation is still missing. 

In this work, we  focus on a simpler problem, yet of great relevance for a broad spectrum
of applications, namely the advection-diffusion-reaction (ADR) equation.
For simplicity, the paper is confined to the case of one spatial dimension.

The paper is organized as follows. In Sec.~\ref{sec:II}, we present the ADR equation and its most salient features. 
Contrary to quantum mechanics, which exhibits a linear and unitary dynamics, the ADR system 
is generally non-linear and non-unitary. 
Hence, to embed the problem on a quantum computer, we need to transform the 
ADR equations accordingly. In order to deal with nonlinearity
we introduce the Carleman linearization procedure and apply it to the ADR equation. 
In Sec.~\ref{sec:III}, we explore the convergence of the Carleman procedure while varying the main parameters,
primarily the Peclet number, namely the ratio of convective versus diffusive transport.

The decomposition of the resulting matrix in terms of Pauli gates is then presented in Sec.~\ref{sec:IV}, where we show that the number of Pauli gates grows exponentially with the number of qubits, highlighting the need for a nontrivial encoding of the matrix in the quantum circuit. This is the topic of the next section.

The non-unitarity of the system can be tackled by exploiting the block encoding of the 
dynamics onto a larger Hilbert space, by using a number of ancilla qubits. 
This method is introduced and analyzed in Sec.~\ref{sec:V}, where 
we write the quantum circuit in explicit form. However, it is shown that the success probability of such circuit needs to be significantly increased in order to ensure the viability of the algorithm on a quantum computer.

\section{Advection-diffusion-reaction equation\label{sec:II}}

In this paper, we focus on the the one-dimensional ADR equation with logistic reaction, 
namely:
 \begin{align}\label{eq:ADR}
     \partial_t \phi = D\partial_{xx}^2 \phi - \partial_x(U \phi) - a \phi + b\phi^2 \text{ ,}
 \end{align}
with $D$ being the diffusion coefficient, $a$ and $b$ are positive constants, and $U$ is the velocity field.
The main physical parameters are the Peclet number $\text{Pe}^L=UL/D$ (strength of advection/diffusion),
the advective Damkohler number $\text{Da}_A^L = \frac{U}{aL}$ (Advection/Reaction) and the 
diffusive Damkohler  number $\text{Da}_D^L= \frac{D}{aL^2}$, (where $L$ is the macroscopic scale of the 
problem. High Peclet numbers are typical of fast macroscopic flows, such as those met
 in atmospheric applications, while high Damkohler numbers describe weakly reactive flows in which
chemical time scales are longer than the transport ones. Typical examples are well-stirred reactors, whereas the opposite case is typical of combustion flames \cite{Bird2006}.
Even without entering the enormous literature on the subject, it is clear that the three-parameter
space $\lbrace \text{Pe},\text{Da}_A,\text{Da}_D \rbrace$ covers a very broad spectrum of reactive transport 
phenomena in physics, engineering, and biology \cite{Bird2006}.

The single-site version of the ADR equation $(D=U=0)$ is the logistic equation, which presents a 
stable attractor, $\phi_{-}=0$, and an unstable one $\phi_+ = a/b \equiv C$, $C$ 
being the so-called carrying capacity of the system. 

Any initial value $\phi_0$ below $C$ decays to zero, and is hence the case of physical interest, whereas any value
above it runs away to infinity in a finite time (finite-time singularity). 
The strength of the nonlinearity is given by the dimensionless ratio $R = \frac{\phi_0}{C}$.

In the presence of multiple sites $x$, we define the 
strength of the nonlinearity as $R = \frac{\phi^{\max}}{C}$
with $\phi^{\max} \coloneqq \max[\phi(x, t = 0)]$. 
In particular we use a box function as initial profile, hence $\phi^{\max}$ is simply the height of the box. 

\subsection{Carleman linearization of the ADR equation}

Here we outline the procedure for the Carleman linearization of the 1D ADR equation~\eqref{eq:ADR} with $N$ sites. 

The procedure is readily extended to any polynomial nonlinearity or 
to higher spatial dimensions. We can write Eq.~\eqref{eq:ADR} in matrix form
\begin{align}\label{eq:matrix_form_evolution}
    \dot \phi_i = A_{ij} \phi_j + b \phi_i^2 ,
\end{align}
where we use Einstein's convention of implicit summation over repeated indices. 
The explicit form of the matrix $A$ depends on the velocity field $U$.

The Carleman procedure~\cite{carleman_application_1932,itani_analysis_2022,sanavio_three_2024} consists in considering the nonlinear terms 
$\phi_i\phi_j\equiv \phi_{ij}$ as new independent variables, which turns the
original nonlinear equation into a infinite hierarchy of linear ones.
The dynamics of the additional variables is readily written down as follows: 
\begin{align}
    \dot{\phi}_{ij} = \dot{\phi_i} \phi_j+\phi_i \dot{\phi_j} = A_{ik} \phi_k\phi_j+A_{jk} \phi_i\phi_k + b \phi_{ii}\phi_j+ b \phi_{jj}\phi_i,
\end{align}
that is now linear in second order terms $\phi_{ij}$ but contains a third order term ($\phi_{ijk}$), 
then to be treated  as a new variable. 
Repeating the procedure transforms a nonlinear equation with polynomial nonlinearities 
into an infinite system of linear equations. 
In practice, the system is truncated to obtain an approximation of the original equation 
in terms of a finite linear system. 

To be more specific, our equation can be written as 
\begin{align}
    \dot \phi = A \phi + B \phi^{\otimes 2},
\end{align}
In index notation for a quadratic nonlinearity, it reads
\begin{align}
    \dot \phi_i = A_{ij} \phi_j + B_{ijk} \phi_{jk} \text{ .}
\end{align}
In our particular case the $N \times N^2$ matrix $B$ takes the 
simple 1-sparse form (no summation over $j$)
\begin{align}
    B_{ijk} = b \delta_{ij}\delta_{jk}.
\end{align}
We then introduce the Carleman variables
\begin{align}\label{eq:carl_variables}
    u_n =& \phi^{\otimes n} \text{,}\\
    \bf{u} =& (u_0, u_1, u_2, \dots)^{\text{T}} \text{ ,}
\end{align}
that leads to an infinite system of coupled linear equations. 

To find a numerical solution, we truncate the Carleman embedding at order $K$,
so that: 
\begin{align}\label{eq:carl_variables2}
    \bf{u} =& (u_0, u_1, u_2, \dots, u_K)^{\text{T}} \text{ ,}
\end{align}
and the linearized set of equations becomes
\begin{align}\label{eq:linearized_problem}
    \dot{\bf{u}} = \mathcal{C}\bf{u} \text{ .}
\end{align}
In the case of a quadratic nonlinearity the Carleman matrix $\mathcal{C}$  
takes the simple structure
\begin{align}\label{eq:carl_matrix}
\mathcal{C}  = 
    \scalemath{0.85}{
    \begin{bmatrix}
A & B & 0 & \cdots & 0\\
0 & A \otimes \mathbb{1} + \mathbb{1} \otimes A & B \otimes \mathbb{1} + \mathbb{1} \otimes B & \cdots & 0 \\
0 & \ddots & \ddots & \cdots & 0\\
0 & \cdots  & \sum_{i = 0}^k \mathbb{1}^{\otimes i} \otimes A \otimes \mathbb{1}^{\otimes (k - i)} & \sum_{i = 0}^k \mathbb{1}^{\otimes i} \otimes B \otimes \mathbb{1}^{\otimes (k - i)} & \cdots \\
0 & \ddots & \ddots & \ddots & \sum_{i = 0}^K \mathbb{1}^{\otimes i} \otimes A \otimes \mathbb{1}^{\otimes (K - i)}
\end{bmatrix}
}
\end{align}

In the above, $\mathbb{1}$ is the $N \times N$ identity matrix, and in 
our notation $\mathbb{1}^{\otimes 0}\otimes A = A$. 

Next, we solve the truncated system with an Euler's scheme, yielding: 
\begin{align}\label{eq:euler_carl}
    \mathbf{u}(t + \Delta t) = (\mathbb{1} + \Delta t \mathcal{C}) \mathbf{u}(t) \text{ .}
\end{align}

In the next section, we present the numerical results for the ADR at 
different Peclet numbers, showing that the convergence of the linearized system to the 
solution of the nonlinear equation is still governed by the reaction term even at high Peclet numbers. 

This implies an exponential convergence with the Carleman truncation level, a necessary 
(but not sufficient) requirement for a resource-efficient implementation 
of the Carleman-based quantum algorithm. In fact, the size of the linearized 
problem grows as $\sim N^K$ for large $N$, hence becoming rapidly unviable 
at growing $K$ even on quantum computers.

\section{Numerical results}\label{sec:III}
\subsection{Constant velocity field $U(x)=U$}

By using a finite-difference scheme and the Euler's forward method, the ADR 
equation takes the form
\begin{align}\label{eq:ADR_fe_costant_c}
\dot{\phi}_j = \frac{D}{\Delta x^2} (\phi_{j-1}^t -2 \phi_j + \phi_{j+1}) - \frac{U}{2 \Delta x} (\phi_{j + 1} - \phi_{j-1}) 
            - a \phi_j + b (\phi_j)^2, \;j=1,\dots,N \text{ ,}
\end{align}
with periodic boundary condition  $\phi(0,t)= \phi(L,t)$.
We choose units such that $\Delta x = D = 1$, that is, distances are measured 
in terms of lattice units and time is measured in units of $\Delta x^2/D$ (cell diffusion time). 

The matrix $A$ of Eq.~\eqref{eq:matrix_form_evolution} writes as follows:

\begin{align}
    A_{ij}=(-2D-a)\delta_{ij}+\left(D-\frac{U}{2}\right)\delta_{i,j+1}+\left(D+\frac{U}{2}\right)\delta_{i,j-1}.
\end{align}

We study the solution of the ADR equation at various values of the Peclet number 
and the logistic parameter $R$ to show that the convergence to the solution of the 
nonlinear equation is mainly determined by the logistic part, with a weak dependence 
on the cell Peclet number $U \Delta x/D$, even for values $\text{Pe} > 2$, where 
we observe numerical fluctuations with the finite difference scheme. 

In all the numerical calculations, we use $N = 20$ lattice sites and 1000 timesteps 
with a time step  $\Delta t = 0.01$ and a box function
\begin{align}
    \phi_j=\begin{cases}
        \phi^{\max} \quad&\text{for }j=\frac{N}{2}+w\\
        0 \quad&\text{otherwise}
    \end{cases}
\end{align}

\noindent with height $\phi^{\max} = 1$ and width $w=5$ as initial condition. 
The Carleman embedding is truncated at $K = 5$. 

At first, we fix the strength of the nonlinearity at $R = 0.6$ by setting $a = 1$ and $b = 0.6$ 
(notice that for the same $R$ but $a, b \ll 1$ the equation is dominated by the 
advection-diffusion part, and the non-linearity is negligible) and vary the Peclet number.  

\begin{figure}[!h]\centering
  \begin{subfigure}{.5\textwidth}
  \subcaption{}
    \centering
    \includegraphics[width=\linewidth]{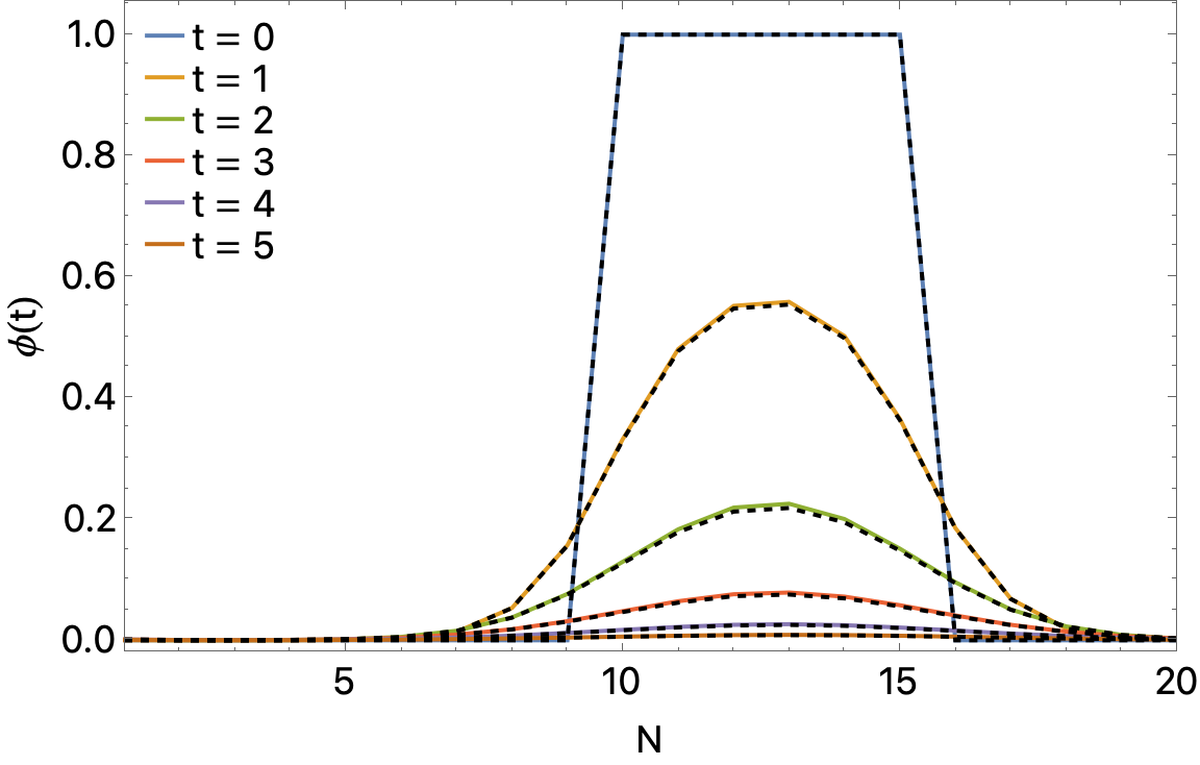}
  \end{subfigure}%
\begin{subfigure}{.5\textwidth}
  \centering
  \subcaption{}
  \includegraphics[width=\linewidth]{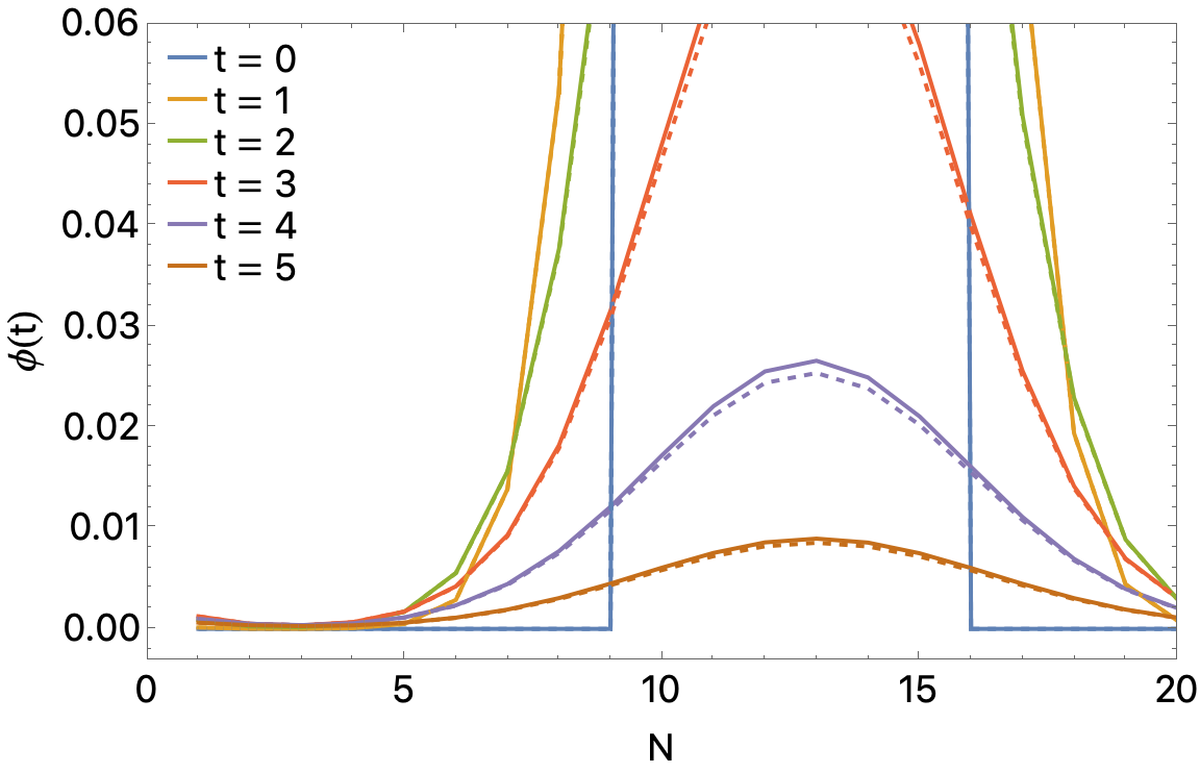}
\end{subfigure}\\
\begin{subfigure}{.5\textwidth}
  \centering
  \subcaption{}
  \includegraphics[width=\linewidth]{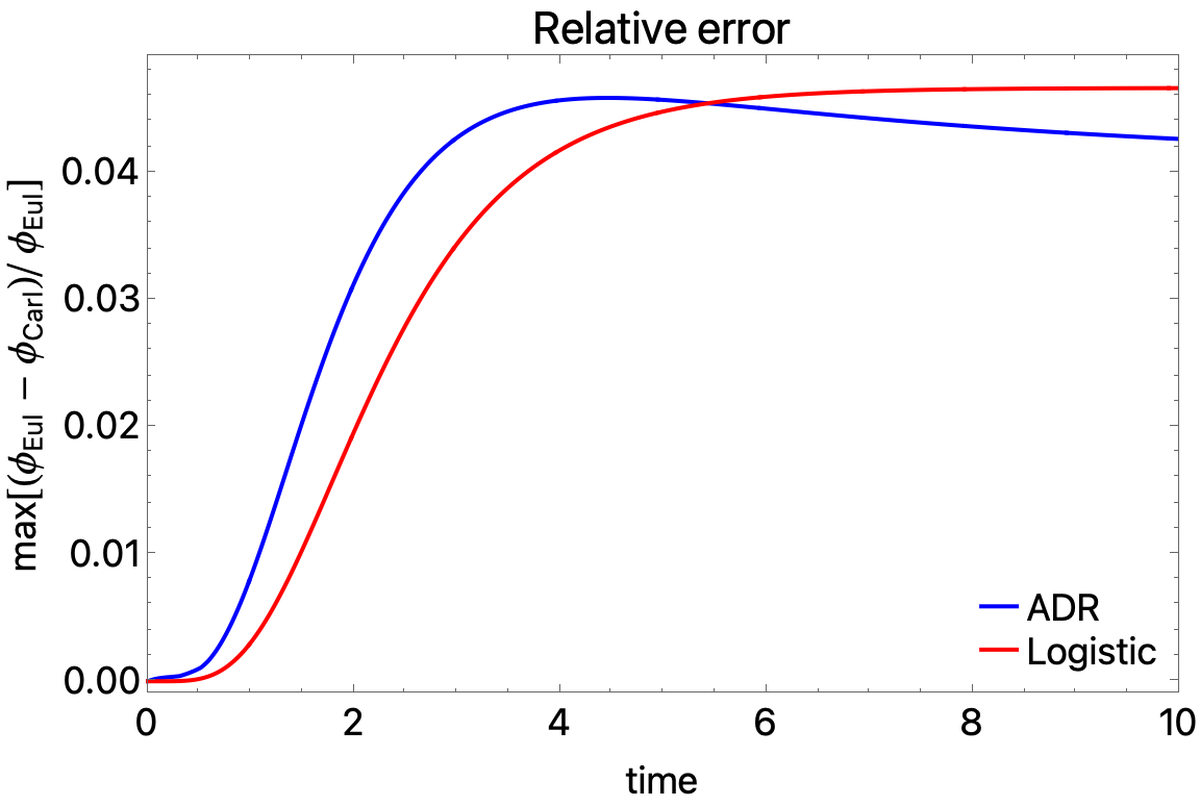}
\end{subfigure}%
\begin{subfigure}{.5\textwidth}
  \centering
  \subcaption{}
  \includegraphics[width=\linewidth]{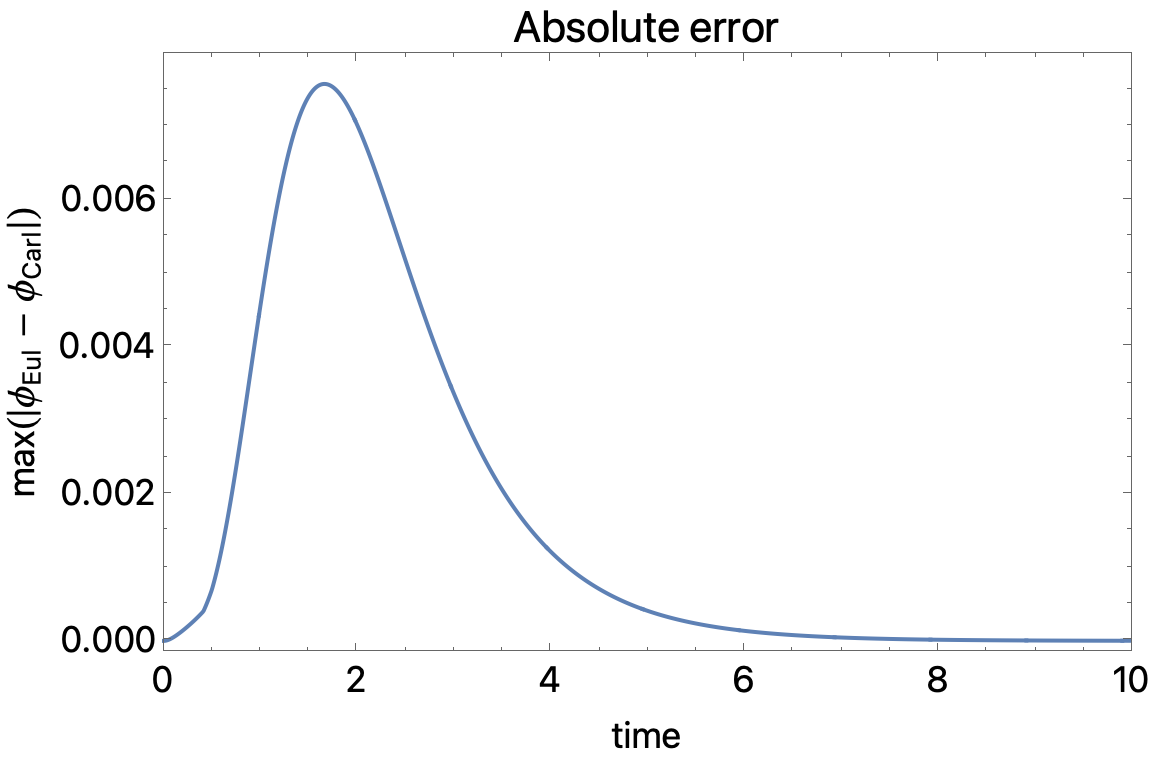}
\end{subfigure}
\caption{\label{fig:peclet_0_1_R_0_6} Solution to the ADR equation with $\Delta t = 0.01$, $N_\text{steps} = 1000$, $a = 1$, $b = 0.6$ ($R = 0.6$), and $\text{Pe} = 0.1$. (a) Comparison between solution with Euler's method and Euler's with Carleman (dashed); (b) zoomed-in version; (c) relative error of the Carleman approximation for ADR and comparison with the equivalent logistic; (d) absolute error of the Carleman approximation.}
\end{figure}

In Fig.\ref{fig:peclet_0_1_R_0_6} we show the results at various 
times for $\text{Pe} = 0.1$, where the dynamic is dominated by diffusion. 
We compare $\phi_\text{Carl}(t)$, the solution to the linearized Carleman system 
with $K = 5$ (colored lines in the plot), and $\phi_\text{Eul}(t)$ the solution of the 
nonlinear equation using Euler's method (dashed black lines). 

In the bottom left panel, we compare the relative error defined as 
\begin{align}
    \text{max}|\phi_\text{Eul}(t) - \phi_\text{Carl}(t)|/|\phi_\text{Eul}(t)| ,
\end{align}
to the error given by the truncation to fifth order of the Carleman linearization 
of the corresponding logistic equation (red line). 

The logistic equation has the exact solution 
\begin{align}
\label{EXA}
\phi_\text{log}(t) = \phi_0 \frac{e^{-at}}{1- (1-e^{-at})} \text{ ,}
\end{align} 
and the Carleman linearization can be applied explicitly resulting in  
\begin{align}\label{eq:logistic_carleman}
\phi_\text{log}^{(K)}(t)=\phi_0e^{-at}\sum_{k=0}^K\bigg{[}R(1-e^{-at})\bigg{]}^k,
\end{align}
for the truncation at order $K$. The error for the logistic is then computed as 
\begin{align}\label{eq:error_logistic}
    \text{max}|\phi_\text{log}(t) - \phi_\text{log}^{(5)}(t)|/|\phi_\text{log}(t)| \text{ .}
\end{align}

For the sake of completeness, we mention that we did the calculation also for the cases $\text{Pe}=2,3$ with no qualitative differences w.r.t. the lower $\text{Pe}$ cases.
Interestingly, for $\text{Pe}>2$, where the Euler scheme leads to an unstable solution, the Carleman linearization still converges to the classical (unstable) solution.

\begin{figure}[!h]\centering
\begin{subfigure}{.5\textwidth}
  \subcaption{}
    \centering
    \includegraphics[width=\linewidth]{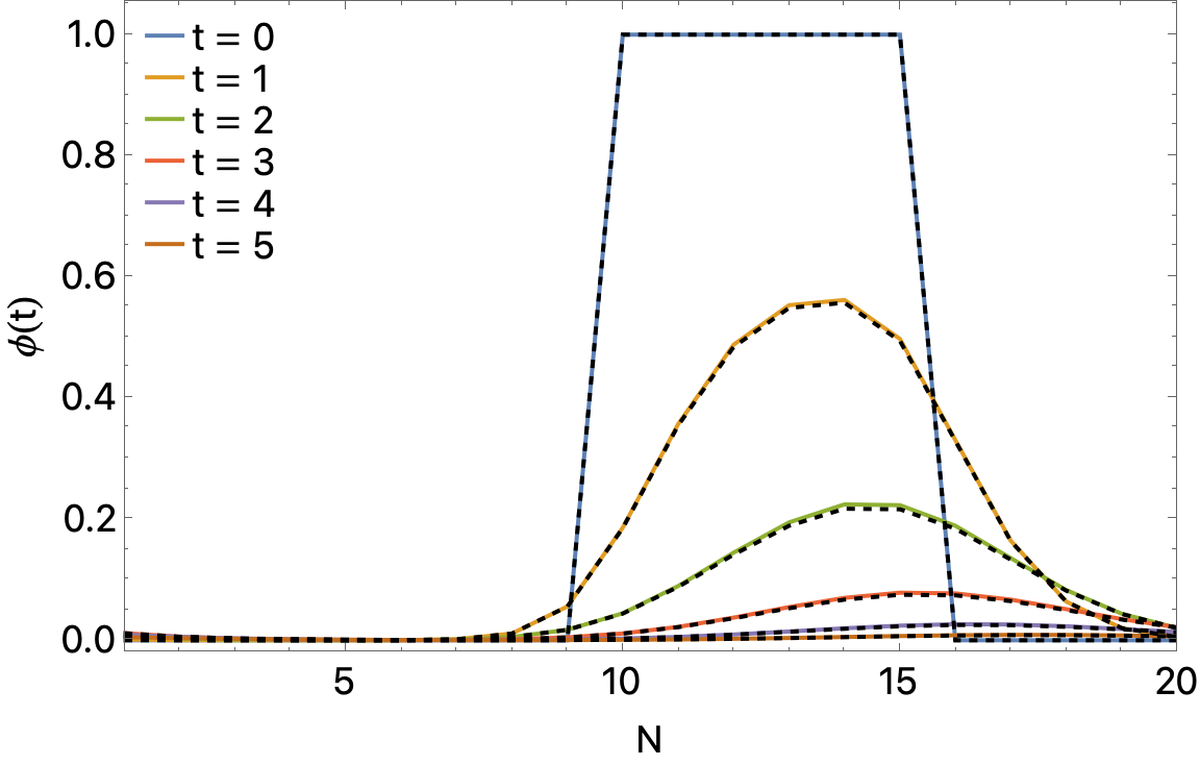}
  \end{subfigure}%
\begin{subfigure}{.5\textwidth}   \subcaption{}
  \centering
  \includegraphics[width=\linewidth]{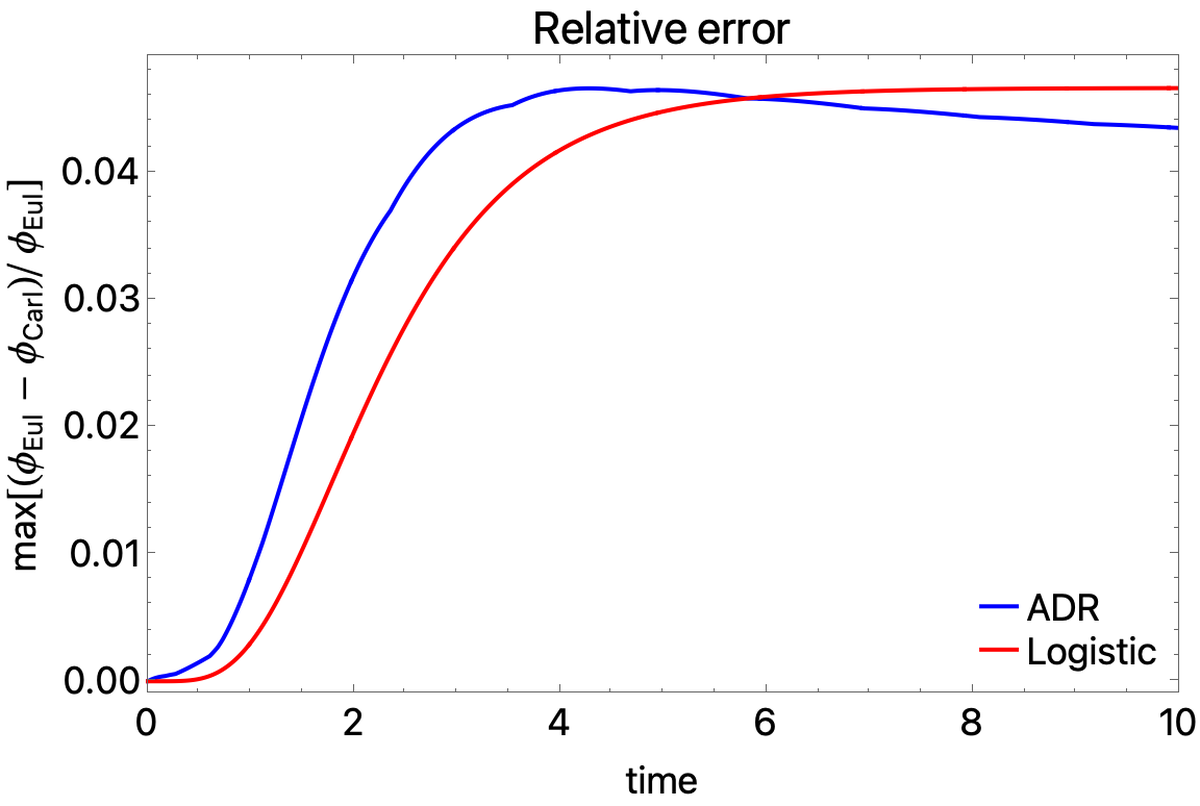}
\end{subfigure}\\
\begin{subfigure}{.5\textwidth}   \subcaption{}
  \centering
  \includegraphics[width=\linewidth]{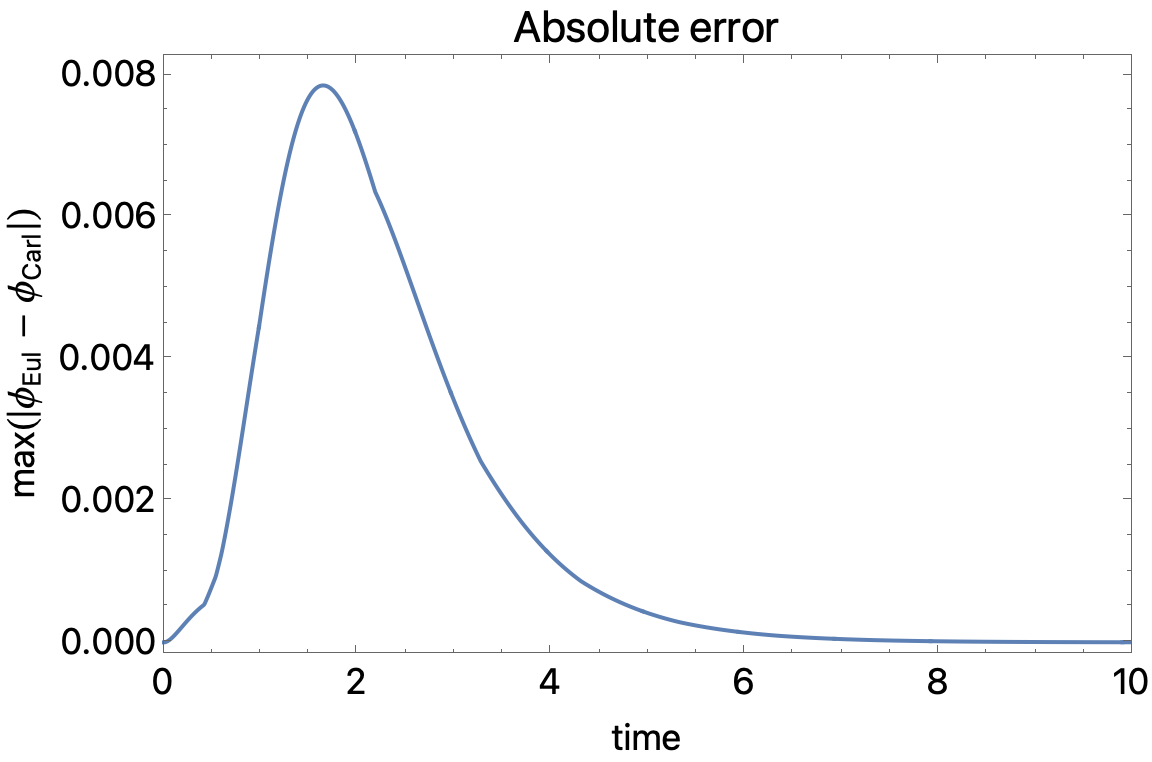}
\end{subfigure}%
\begin{subfigure}{.5\textwidth}   \subcaption{}
  \centering
  \includegraphics[width=\linewidth]{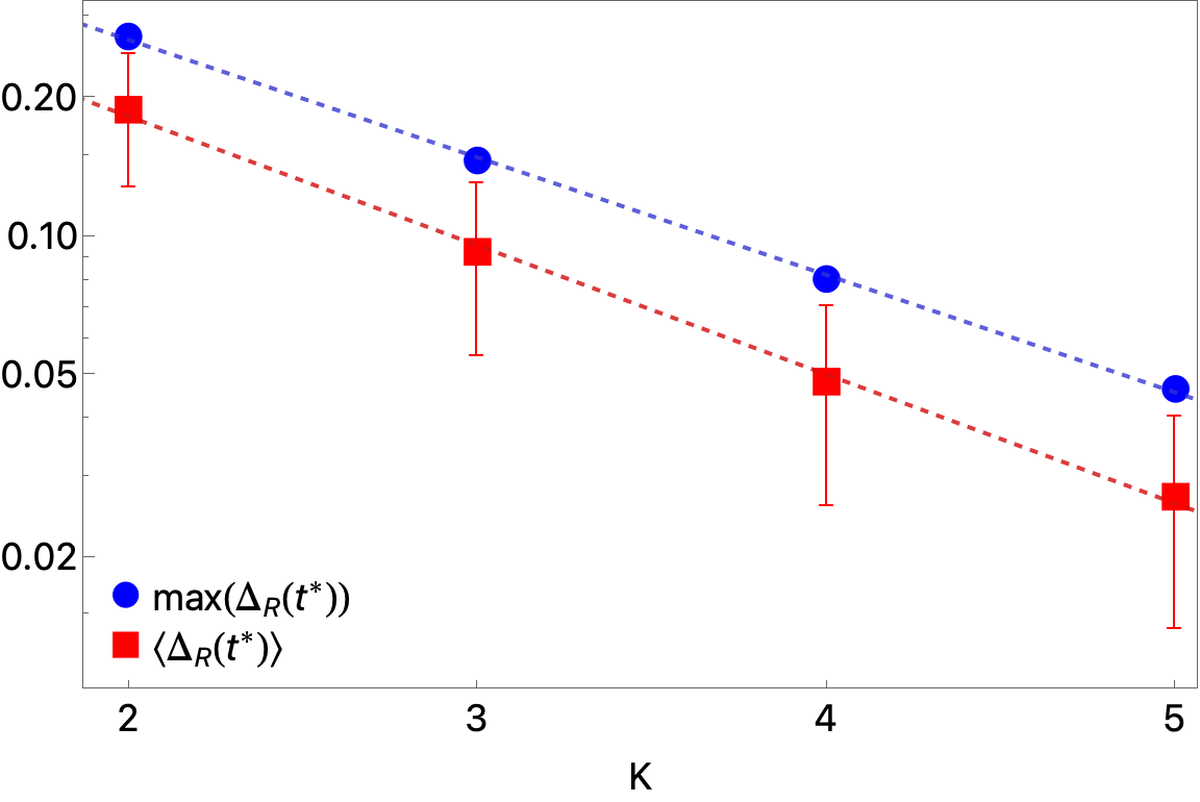}
\end{subfigure}
\caption{\label{fig:peclet_1_R_0_6} Solution to the ADR equation with $\Delta t = 0.01$, $N_\text{steps} = 1000$, $a = 1$, $b = 0.6$ ($R = 0.6$), and $\text{Pe} = 1$. (a) Comparison between solution with Euler's method and Euler's with Carleman (dashed); (b) relative error of the Carleman approximation for ADR and comparison with the equivalent logistic; (c) absolute error of the Carleman; (d) exponential convergence  for increasing Carleman truncation level $K$calculated at $t=t^*$, the maximum value of the relative error $\max \Delta_R(t^*)$ (blue circle), and 
the mean value averaged over the lattice sites $\langle \Delta_R(t^*)\rangle$ (red square).}
\end{figure}

Since the error depends mostly on the logistic part of the equation, for the values 
considered here we expect an exponential convergence to the solution as a function 
of the truncation order $K$. This is indeed the case, as shown in Fig.\ref{fig:peclet_1_R_0_6}(d), 
where we present a logarithmic plot of the maximum of the relative difference, $\max(\Delta_R(t^*))=\max_t|\phi_\text{Eul}(t) - \phi_\text{Carl}(t)|/\phi_\text{Eul}(t)$, as a function of $K$, where $t^*$ is the time when the maximum is reached, together with the plot of the relative error averaged over the lattice sites $\langle\Delta_R$, at $t=t*$.

\begin{figure}[!h]\centering
  \begin{subfigure}{.5\textwidth}   \subcaption{}
    \centering
    \includegraphics[height=0.21\textheight]{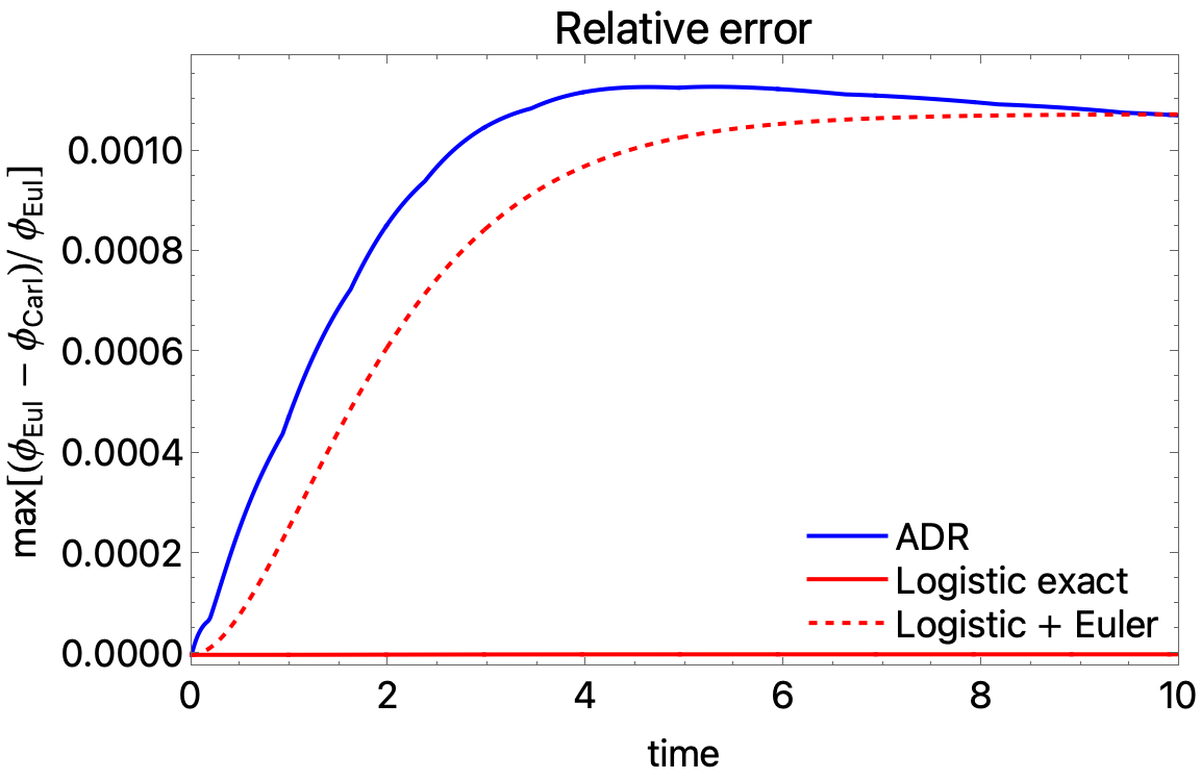}
  \end{subfigure}%
\begin{subfigure}{.5\textwidth}   \subcaption{}
  \centering
  \includegraphics[height=0.21\textheight]{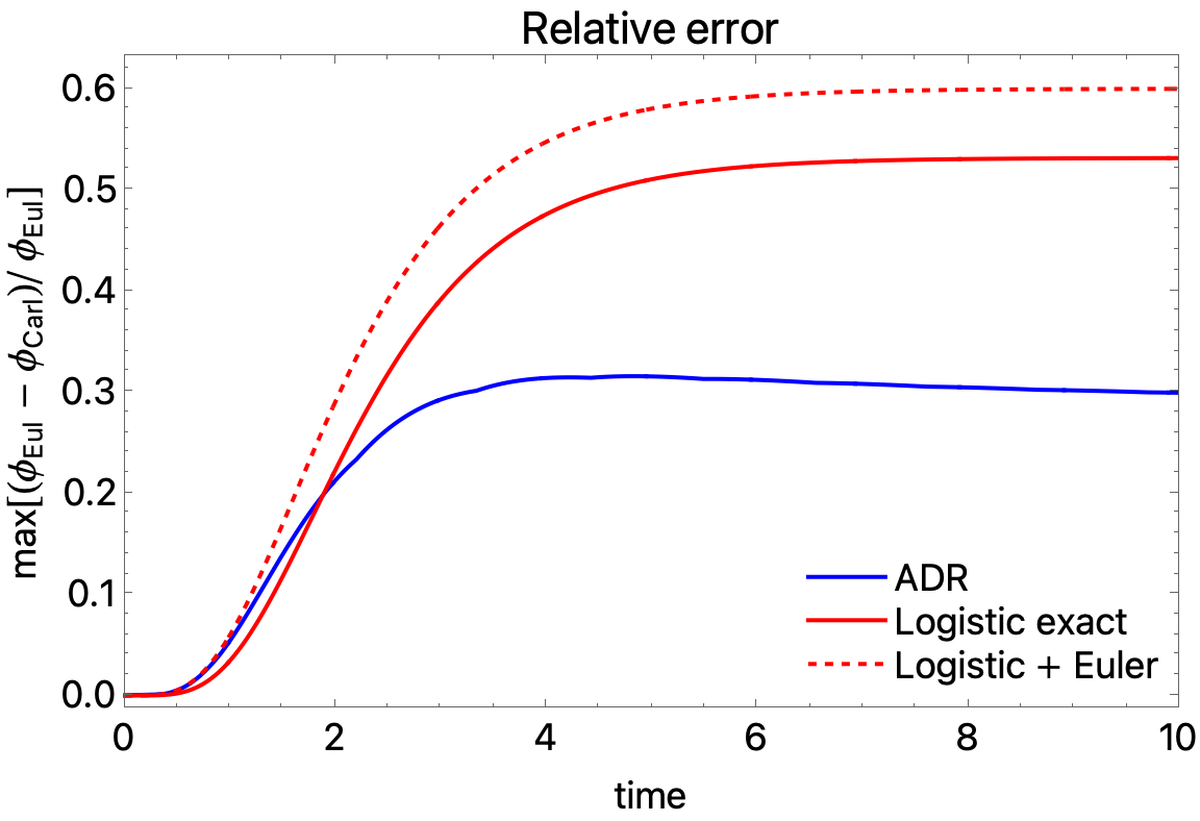}
\end{subfigure}\\
\caption{\label{fig:rel_error_R} Relative error between the solution of the ADR nonlinear equation with $\text{Pe} = 1$ and the linearized one using Carleman at order 5. (a) Weak nonlinearity $R=0.1$, (b) strong nonlinearity $R = 0.9$. For comparison, we show the error on Carleman applied to the logistic using the exact expressions in Eqs.\ref{EXA} and \ref{eq:logistic_carleman} (solid red line) and using Euler's method (dashed red line). Notice that for strong nonlinearities, the advection-diffusion term helps the convergence of the Carleman approximation as it diminishes the effect of the nonlinear part.}
\end{figure}

Similar results are obtained by fixing the velocity (hence, the Peclet number), $U = 1$, 
with $a = 1$, and vary the value of $b$ to study different $R$. 
In Fig.\ref{fig:rel_error_R} we show the relative error for $R = 0.1$, where the effect of 
the nonlinearity is small. For $R = 0.9$, Carleman converges more slowly and for $R = 0.1$ the 
Carleman error is obviously small but much larger than what is expected for the logistic. 

This is in part due to the fact that for ADR we obtained the Carleman solution 
and the solution to the nonlinear equation using Euler's method, while for the logistic 
we compared the exact expressions with Eq.\ref{eq:logistic_carleman}. 

By using Euler's method for the logistic and the corresponding linearized problem, a 
comparable error is observed (dashed red line in the figure). 
Notice that for $R > 1$ the logistic diverges in finite time for $a, b > 0$, an interesting 
instability whose study goes beyond the scope of the present work.

\subsection{Non-uniform velocity field $U(x)$}
\label{sec:velocity_field}

The case of uniform velocity is highly idealized, since in most cases of practical interest 
the velocity field ("winds") changes in both space and time.

Here we show numerically that the Carleman embedding for ADR can also deal 
with a non-uniform velocity field $U = U(x)$. 

\begin{figure}[!h]\centering
\begin{subfigure}{.5\textwidth}   \subcaption{}
  \centering
  \includegraphics[height=0.21\textheight]{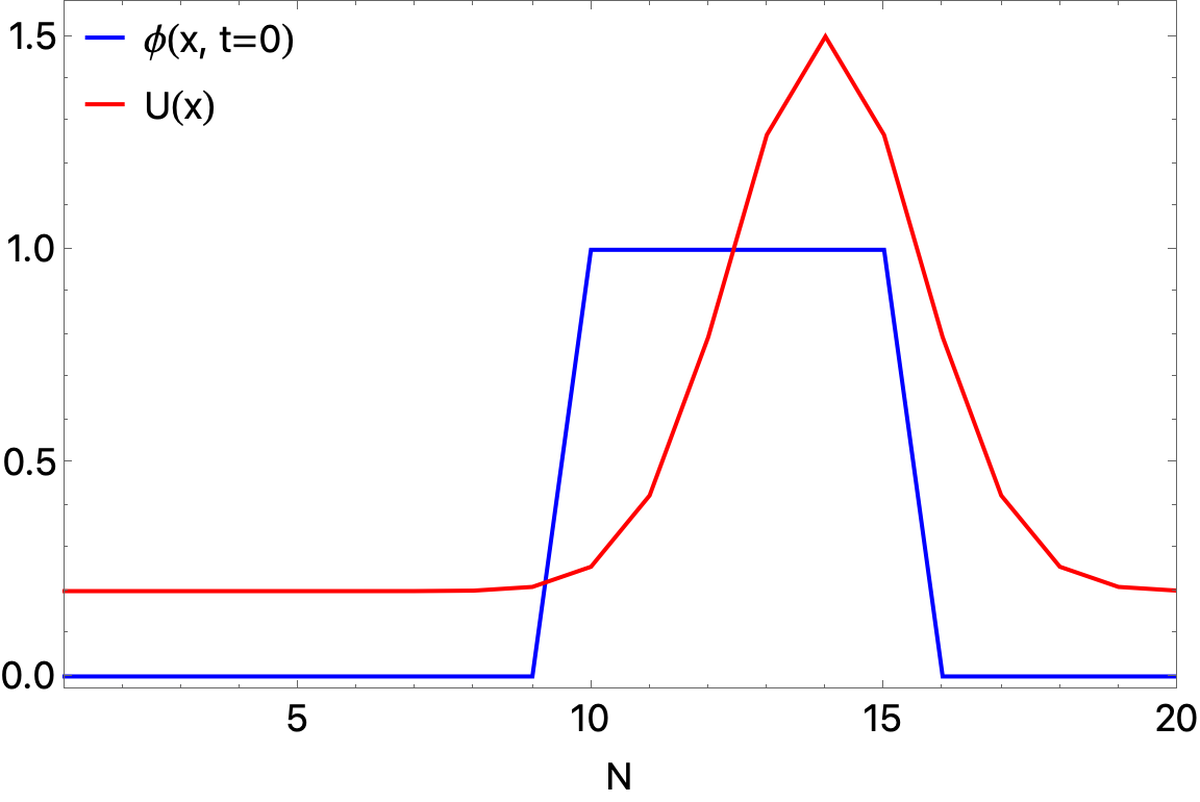}
\end{subfigure}%
  \begin{subfigure}{.5\textwidth}   \subcaption{}
    \centering
    \includegraphics[height=0.21\textheight]{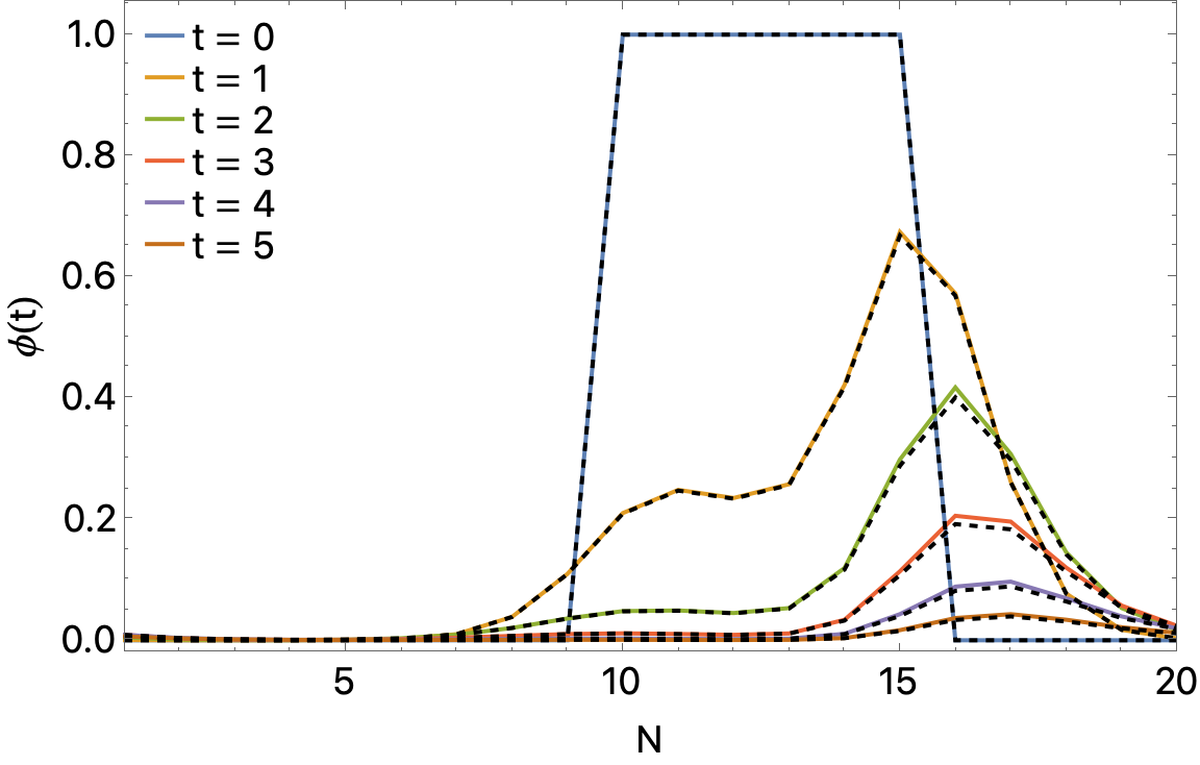}
  \end{subfigure}\\
\begin{subfigure}{.5\textwidth}   \subcaption{}
  \centering
  \includegraphics[height=0.21\textheight]{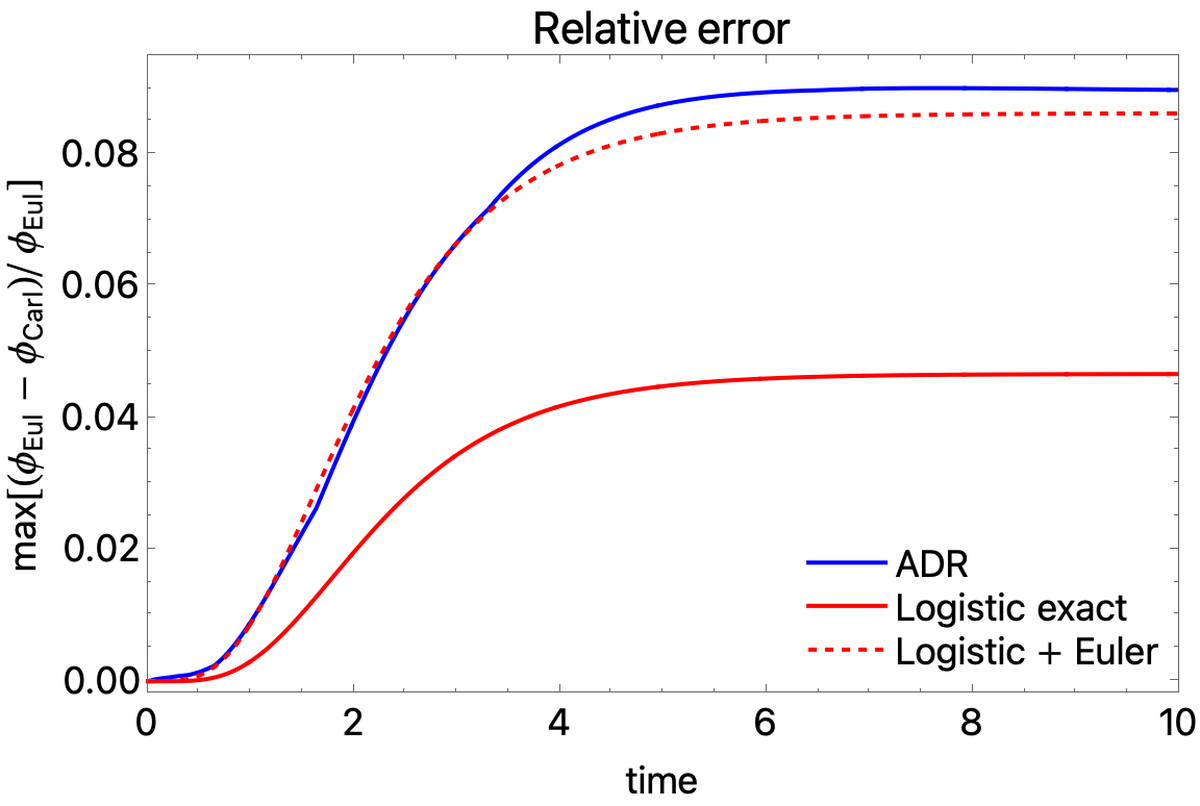}
\end{subfigure}%
\begin{subfigure}{.5\textwidth}   \subcaption{}
  \centering
  \includegraphics[height=0.21\textheight]{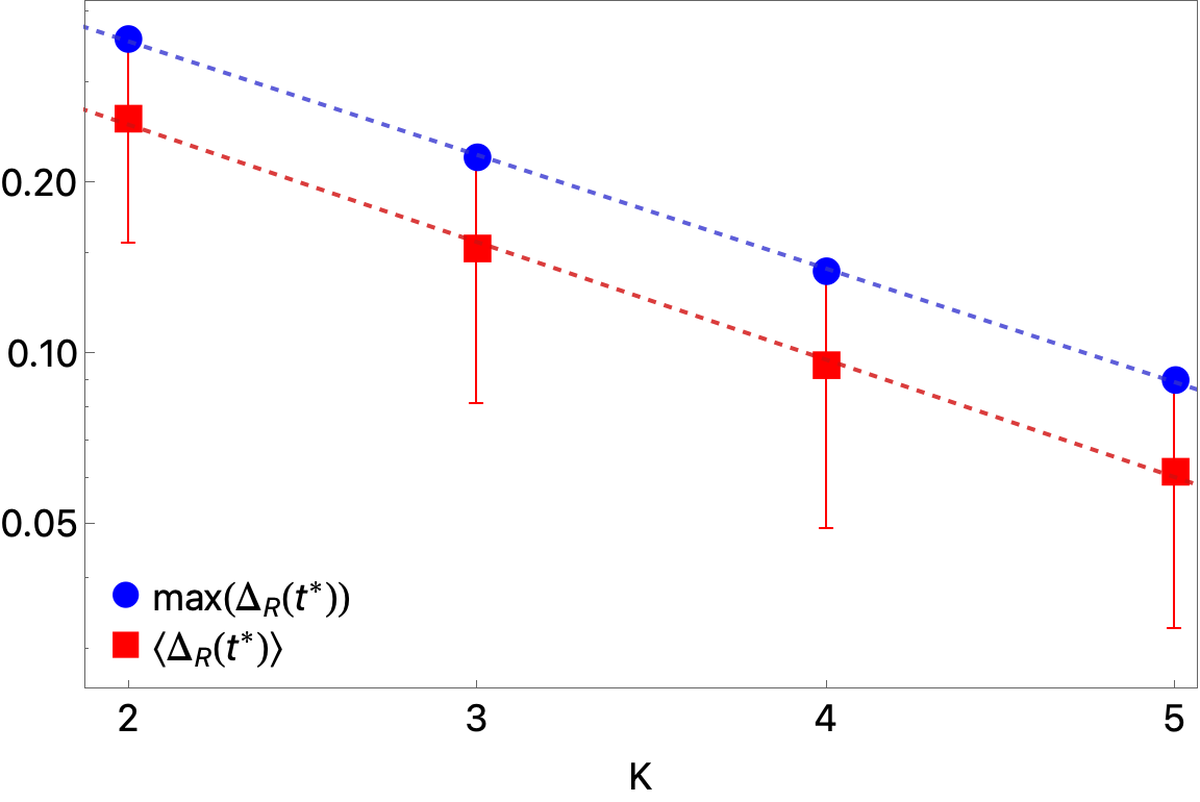}
\end{subfigure}
\caption{\label{fig:vel_field} Solution to the ADR equation with Euler's method with $\Delta t = 0.01$, $N_\text{steps} = 1000$, $a = 1$, $b = 0.6$ ($R = 0.6$), and a Gaussian velocity profile . (a) Comparison between the solution of the Carleman linearized system and the nonlinear equation. (b) Relative error of the Carleman linearization, the maximum is reached at $t=t^*$. (c) Absolute error of Carleman linearization. (d) Convergence of the maximum relative error $\max\Delta_R(t^*)$ and mean relative error $\langle\Delta_R(t^*)\rangle$ in logarithmic scale.}
\end{figure}

Using a central difference scheme, we solve the system:
\begin{align}
    \phi_j^{t + \Delta t} =& \phi_j^t + \Delta t \big{[}\frac{D}{\Delta x^2} (\phi_{j-1}^t -2 \phi_j^t + \phi_{j+1}^t) - \frac{U_j}{2 \Delta x} (\phi_{j + 1}^t - \phi_{j-1}^t) \nonumber
            \\
            &- \left(a + \frac{U_{j + 1}}{2 \Delta x} - \frac{U_{j - 1}}{2 \Delta x}\right)  \phi_j^t + b (\phi_j^t)^2\big{]}\text{ .}
\end{align}
In Fig.\ref{fig:vel_field} we present the results for a Gaussian velocity profile, as shown in the top left panel. 
We use $R = 0.6$ and a Carleman order $K = 5$ as above. 
Even in this case, this truncation level returns a reasonable approximation of 
the nonlinear problem, with a maximal relative error $<0.1$, due mostly to the reaction term, and therefore converging 
exponentially with increasing truncation order. 

Although not exhaustive, the above results indicate that the Carleman linearization 
provides a robust approach to turn advection-diffusion-reaction problems into a formalism
more amenable to a quantum computer implementation. 

In the next sections, we analyze possible Carleman embeddings of ADR equations into a quantum circuit. 

\section{Quantum algorithm for ADR}\label{sec:IV}

After casting the problem in linear form through the Carleman embedding 
and using a standard discretization scheme, the Euler's method, 
we are left with a linear problem of the form $\mathbf{u} (t+\Delta t) = L \mathbf{u} (t)$ with $L = \mathbb{1} + \Delta t C$. 

Substantial work has been devoted to the development of quantum algorithms for 
solving linear systems~\cite{harrow_quantum_2009, childs_quantum_2017, Dervovic2018}.
\textit{Assuming} the existence of oracles that can efficiently prepare the states 
and the matrices $A$ and $B$, the quantum algorithms are shown to
feature a complexity scaling at most as a polynomial of the logarithm of the variables involved \cite{liu_efficient_2021}.  

However, to the best of our knowledge, none of these previous works provide an explicit form
of such oracles, impairing the practical use of the corresponding algorithms. This is exactly the issue that we tackle in Sec~\ref{sec:V}.

Furthermore, it is worth emphasizing that a  potential speedup of the quantum 
algorithm with respect to the classical one, is not the only goal, as 
another major aim of the quantum algorithm is to harness the dimension 
of the Hilbert space to limit the resources needed for the actual quantum simulation. 
Our aim is to understand if the Carleman procedure can be implemented on a quantum computer with an exponential reduction in the number of resources needed.

First, we look at the structure of the Carleman matrix and evaluate its expansion in terms of Pauli matrices. In fact, when the number of Pauli terms is low, one can implement the corresponding circuit by using a low number of gates exploiting standard tools from Hamiltonian simulation techniques~\cite{whitfield_simulation_2011}.

\begin{figure}[!h]
    \centering \includegraphics[width=0.7\linewidth]{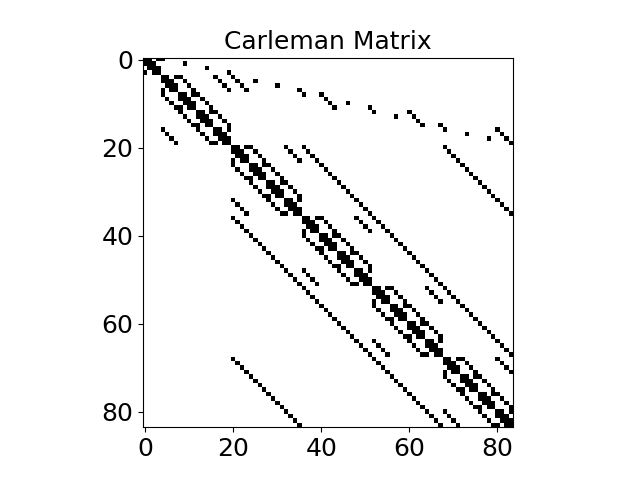}
    \caption{\label{fig:matrix_N4}Structure of the Carleman matrix for $N = 4$ sites and a Carleman truncation level $K = 3$. The black squares show the position of the nonzero elements.}
\end{figure}

In order to evaluate the difference between a matrix and its decomposition 
at various truncation levels, we use the distance induced by the Frobenius norm 
\begin{align}
    \lVert A \rVert = \sqrt{\sum_{i, j} \abs{A_{ij}}^2}.
\end{align}
In Fig.~\ref{fig:matrix_N4} we show the structure of the Carleman matrix 
for $N = 4$ and $K = 3$, where the black squares represent nonzero elements. 
Despite the Carleman matrix being sparse, its decomposition in terms of Pauli matrices 
still requires an exponential number of Pauli terms as we show below. 

To decompose a matrix $\mathcal{C}$ of size $n\times n$ we need at least $q = \lceil \log_2(n) \rceil$ qubits and the corresponding $2^q \times 2^q$ matrix 
\begin{align}
    \mathcal{C}_p = &\begin{bmatrix}
\mathcal{C}_{(n\times n)} & \mathbb{0}\\
\mathbb{0} & {\mathbb{0}}_{(2^q - n\times 2^q - n)}
\end{bmatrix}
\end{align}
can be decomposed on the basis of tensor products of Pauli matrices 
\begin{align}
    \mathcal{C}_p = \sum_{i_1,\dots,i_{q}}\tilde\alpha_{i_1,\dots,i_{q}} \sigma_{i_1} \otimes \dots \otimes \sigma_{i_q} \equiv \sum_{i=1}^{4^q} \alpha_i \Sigma_i
\end{align}
with $\alpha_{i} \in \mathbb{R}$ if $\mathcal{C}_p$ is hermitian, but complex otherwise. Since we are dealing with a sparse matrix, we do not need all the $4^q$ terms of the decomposition, however, for $N > 4$, we find that the number of terms needed to approximate the matrix seems to scale exponentially.



\begin{figure}[!h]\centering
  \begin{subfigure}{.5\textwidth}   \subcaption{}
    \centering
    \includegraphics[width=0.9\linewidth]{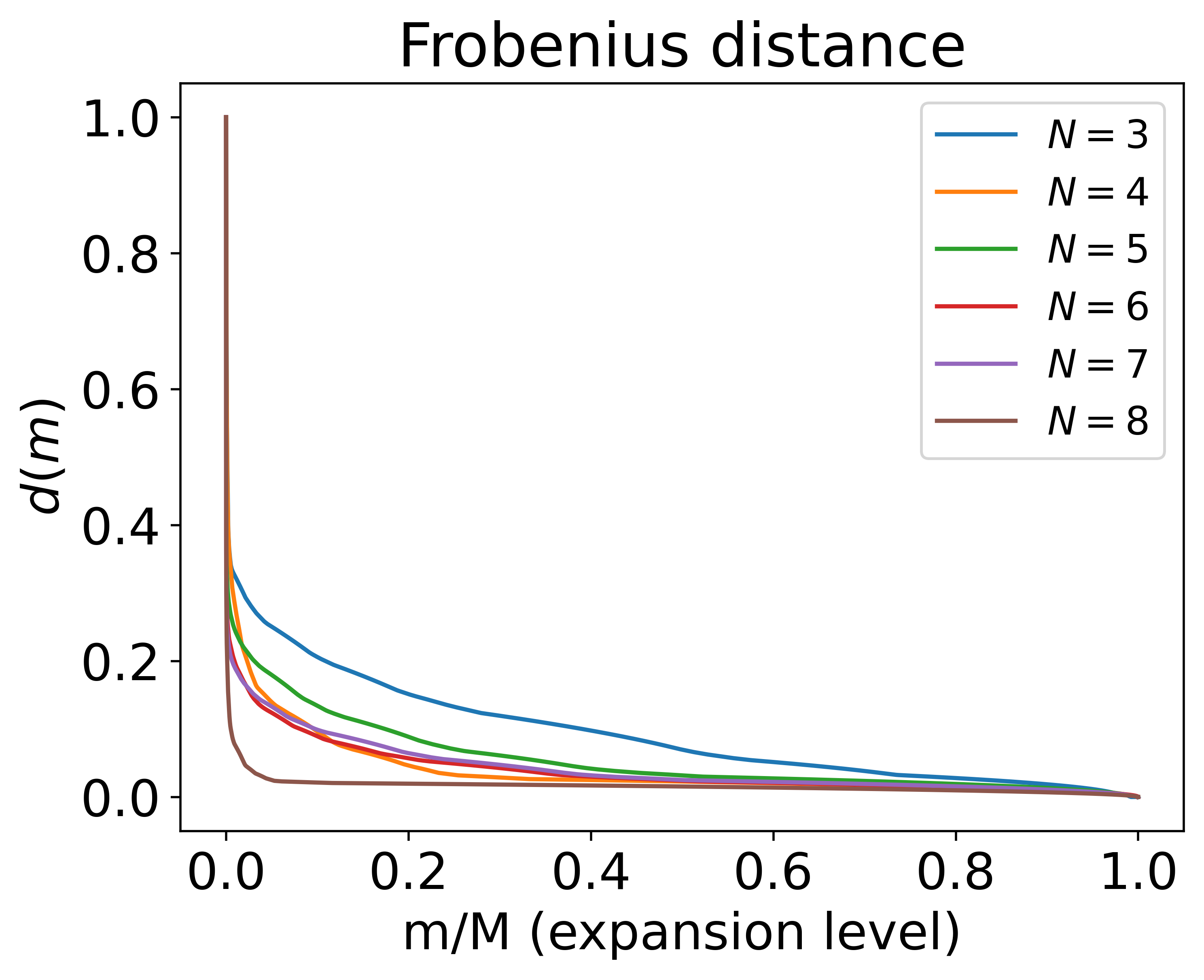}
  \end{subfigure}%
\begin{subfigure}{.5\textwidth}   \subcaption{}
  \centering
  \includegraphics[width=0.9\linewidth]{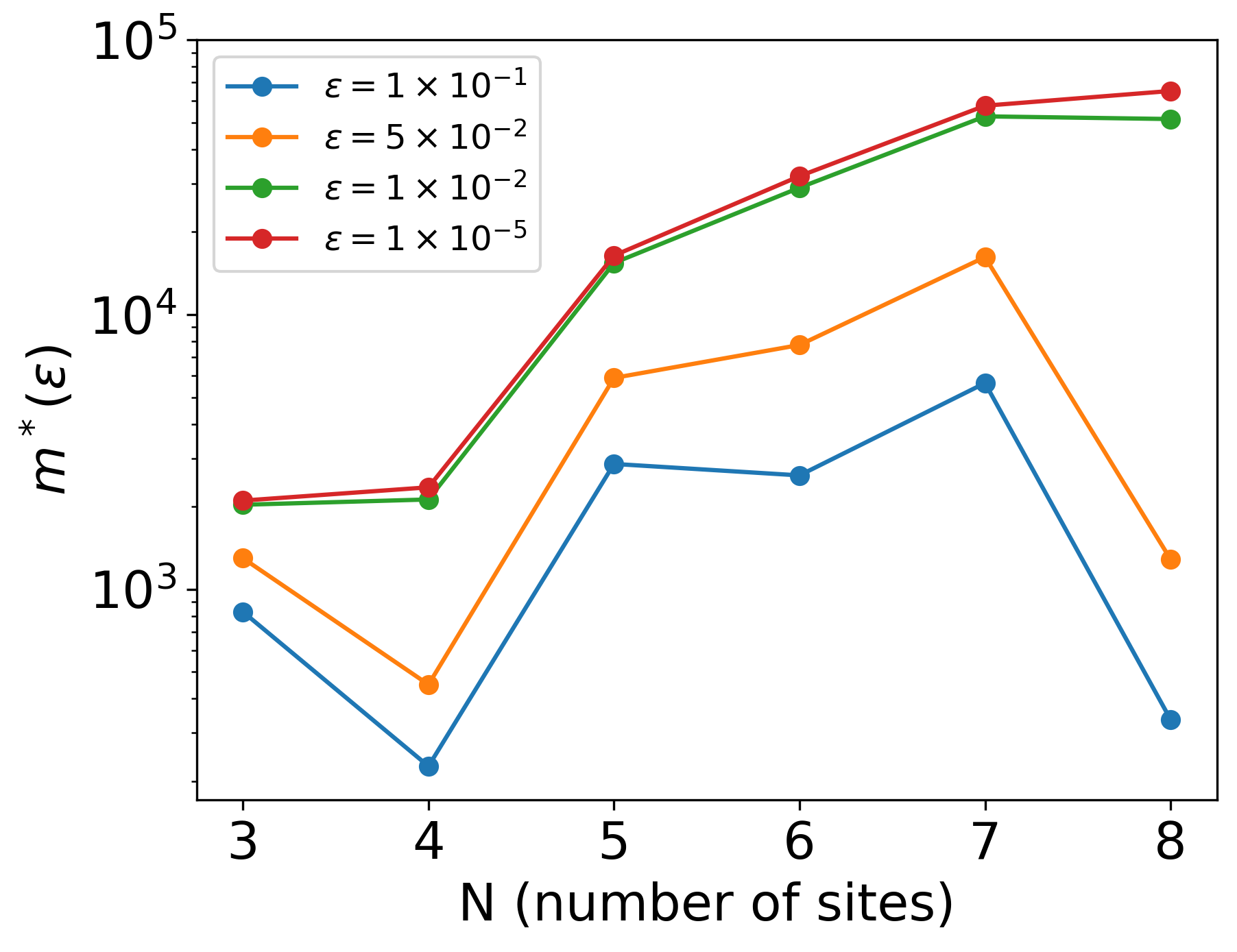}
\end{subfigure}
\caption{\label{fig:carl_nonzero}(a) Normalized Frobenius distance between the matrix $\mathcal{C}_p$ 
and its decomposition in the Pauli basis as a function of the truncation level of the decomposition. 
(b) Dependence of the number of Pauli terms needed to approximate 
$\mathcal{C}_p$ with an error $\epsilon$ on the number of sites, in logarithmic scale.}
\end{figure}

In order to check whether it is possible to find a reasonable approximation 
 of the matrix with only a fewer Pauli terms,  we compute the normalized distance 
 between the Carleman matrix $\mathcal{C}_p$ and the decomposition 
 as a function of the number of terms, $m$, given by
 \begin{align}
     d(m) = ||\mathcal{C}_p - \sum_i^m \alpha_i \Sigma_i||/||\mathcal{C}_p|| ,
 \end{align}
 with the decomposition ordered in terms according to the largest coefficients $\alpha_i$.
 This distance is shown in the left panel of Fig.\ref{fig:carl_nonzero}.  
 In order to compare the behavior for different values of lattice sites $N$, we 
 rescaled the $x$-axis for the total number of nonzero elements $M$ (that is, the x-axis represents 
 the fraction of nonzero elements considered in the decomposition). 
 In the right panel, we show the scaling of the number of terms $m^*$  
 approximating $\mathcal{C}_p$ to within an error $\epsilon$, that is $m^*{:}\, d(m^*) < \epsilon$, where we find that for $N>4$ the number of terms needed to approximate $\mathcal{C}_p$ with an error at least $\epsilon=10^{-2}$ scales exponentially. 

Interestingly, we notice that when the number of sites is a power of two, $N = 2^n$, the Pauli expansion converges much faster to the matrix $\mathcal{C}_p$ (despite the fact that, even when $N = 2^n$, the size of $\mathcal{C}_p$ is not a power of two). As shown in Fig.\ref{fig:carl_nonzero}(b), this in turn implies that for higher values of $\epsilon$ the Carleman matrix can be approximated with few Pauli terms. This advantage seems to disappear when requiring high precision in reconstructing $\mathcal{C}_p$. Because we are interested in a precise approximation of the matrix, and due to the high computational cost of studying the Carleman matrix decomposition at higher powers, we do not explore this behavior further and we leave it to future work.

\begin{figure}[!h]\centering
  \begin{subfigure}{.5\textwidth}   \subcaption{}
    \centering
    \includegraphics[width=0.9\linewidth]{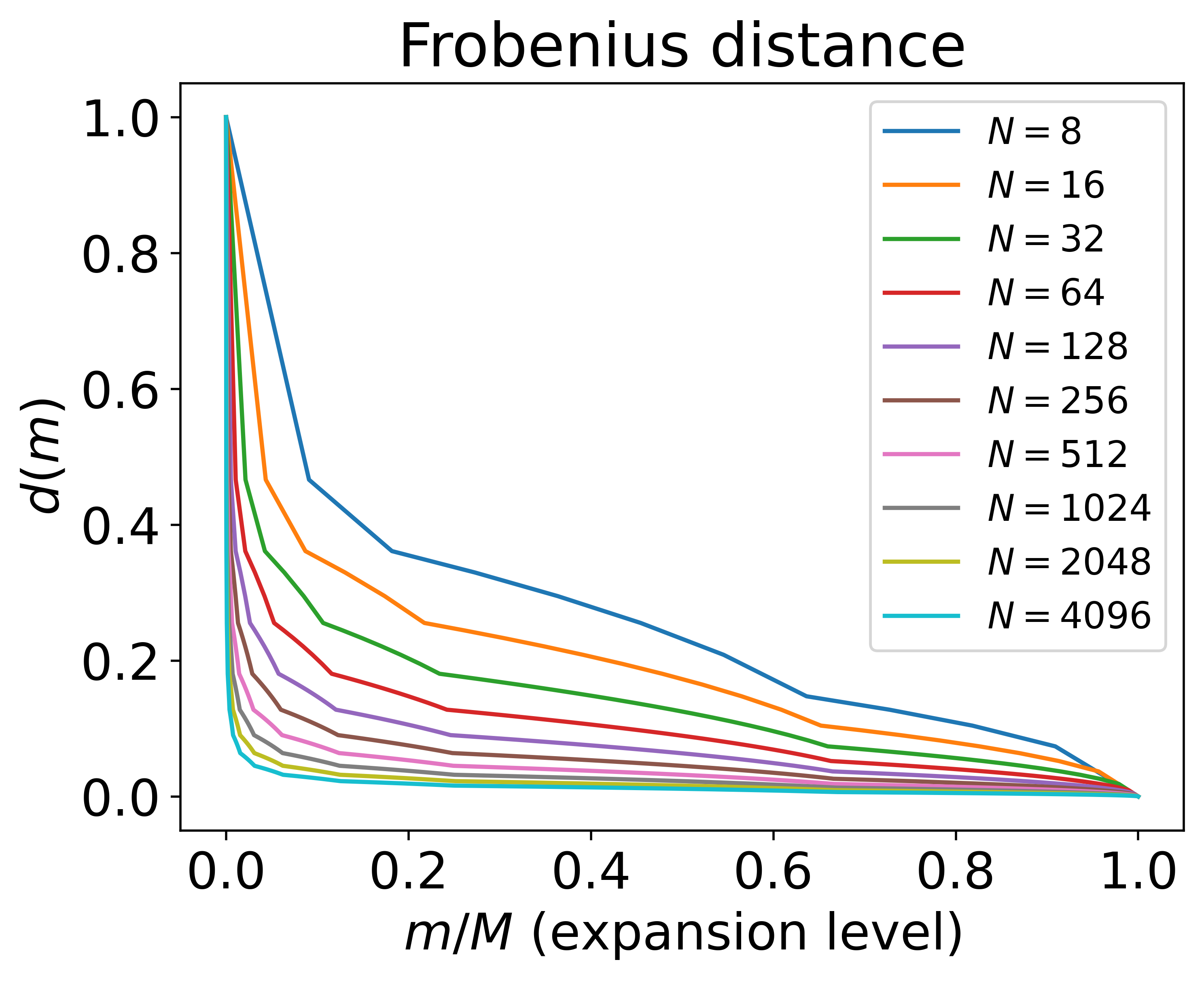}
  \end{subfigure}%
\begin{subfigure}{.5\textwidth}   \subcaption{}
  \centering
  \includegraphics[width=0.9\linewidth]{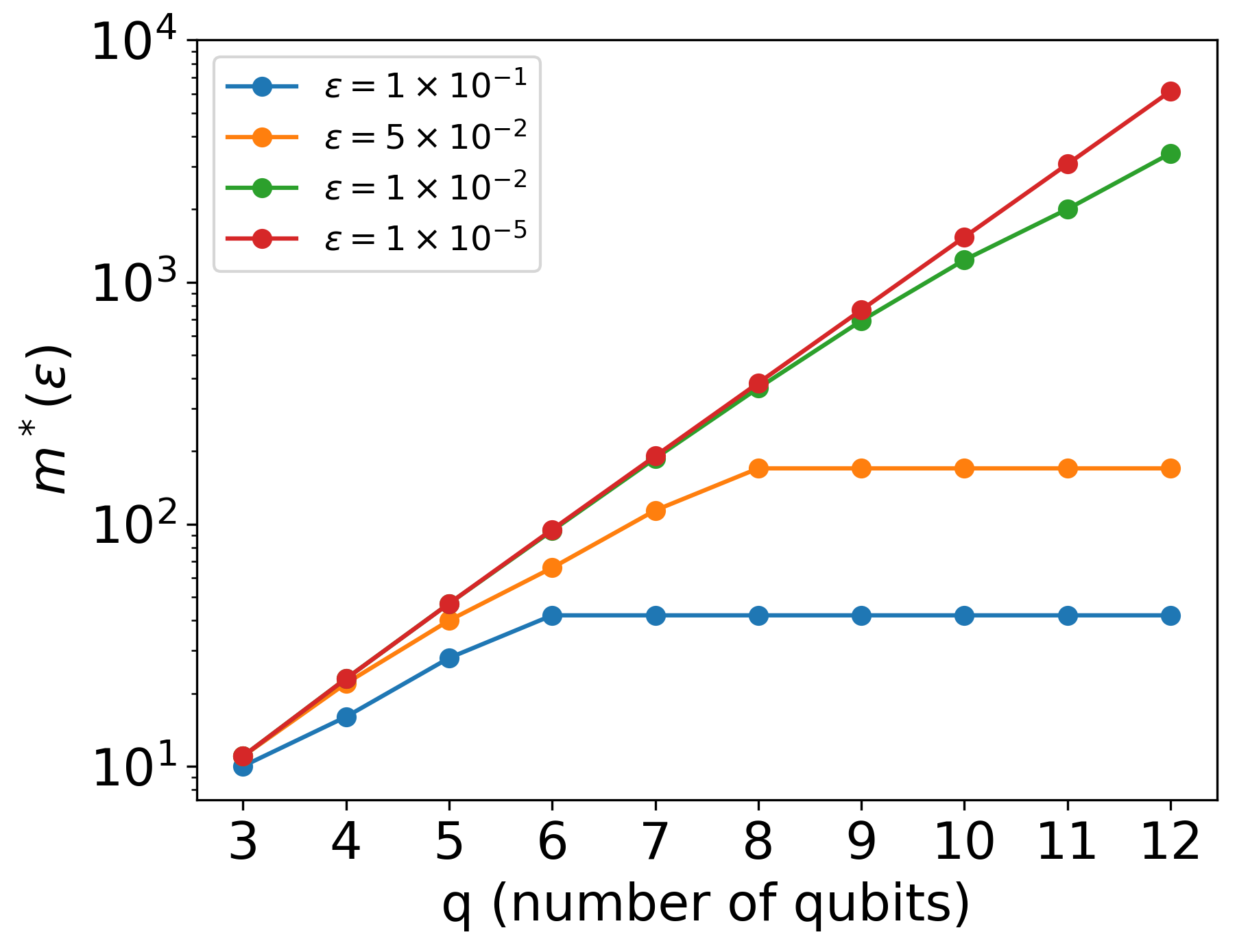}
\end{subfigure}
\caption{\label{fig:a_distance}(a) Frobenius distance between the matrix $A$ with its 
decomposition in the Pauli basis as a function of the truncation level of the decomposition. 
(b) Exponential dependence of the number of Pauli terms needed to approximate $A$ 
with an error $\epsilon$ as a function of the number of qubits needed to encode the matrix.}
\end{figure}

 In order to check the behavior in a simpler case, in Fig.\ref{fig:a_distance}(a) we show the decomposition of 
 the matrix A alone, hence ignoring the nonlinearity, for different values of $N = 2^q$ (again we rescale the $x$-axis for a direct 
 comparison of the behavior for different values of $N$). In Fig.\ref{fig:a_distance}(b) we look at how the number of terms $m^*$, needed to satisfy $d(m^*) < \epsilon$, scales with the number $q$ of qubits needed to encode the matrix $A$. 
 We find that for higher values of the cutoff $\epsilon$, $m^*$ seems to grow slowly after an initial exponential increase with the number of qubits. However, for small $\epsilon$ the number of Pauli terms needed increases exponentially with the number of qubits, highlighting the need for a different encoding strategy.

\begin{figure}[!h]
  \centering
\begin{quantikz}
  \lstick{$|\phi\rangle$}            &\gate{\hat A} &\gate[2]{\hat B}&\qw&\qw&\\
  \lstick{$|\phi\rangle^{\otimes 2}$}&\qw            &               &\gate{\hat A\otimes \mathds{1}+\mathds{1}\otimes \hat A}&\qw
\end{quantikz}
    \caption{Schematic view of the quantum circuit that performs Carleman to second order given two input registers encoding the states $\ket{\phi}$ and $\ket{\phi} \otimes \ket{\phi}$. We remind that $A$ and $B$ are not unitary, hence the 
    block $A$ represents the circuit implementing the non-unitary 
    transformation $A$, and similarly for $B$.}
    \label{fig:carl_circuit}
\end{figure}
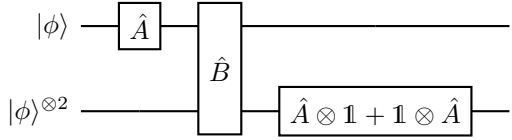

 This shows that there is no trivial way to write an efficient Pauli-based quantum circuit 
 reproducing the Carleman matrix.
 Yet, we can look for methods to efficiently encode the matrices A and B, namely
 to construct a quantum circuit performing  the Carleman embedding,
 according to the scheme shown in Fig.\ref{fig:carl_circuit} for the simple case $K = 2$.

In the following, we discuss a potential candidate for such a strategy, based on the 
notion of block-encoding of sparse matrices.

\section{Block encoding with matrix access oracles}\label{sec:V}

In order to cope with the remaining non-unitarity of the linearised dynamics and 
implement the circuit sketched in Fig.~\ref{fig:carl_circuit} we employ a technique 
called \textit{block-encoding} (BE), which is at the edge of the current technology for 
implementing both unitary~\cite{low_optimal_2017} and non-unitary evolution~\cite{gilyen_quantum_2019}. 

Let's consider a real $s$-sparse matrix $M$ of size $N\times N$, with $N=2^n$, and our goal is to calculate its action on the state $|\psi\rangle$. We restrict the components $M_{ij}$ and the amplitude of the maximum eigenvalue by the value 1.
 The output of our circuit will be the normalized state $M|\psi\rangle/||M|\psi\rangle||$.  If $M$  is not unitary, we can block-encode it on a larger Hilbert space, by writing

\begin{align}\label{eq:block_encoding}
U_M = \begin{pmatrix}
M & *\\
* & *
\end{pmatrix},
\end{align}

\noindent where we labeled the components that are not relevant to us with the symbol `$*$'. 
The particular form of $U_M$ already suggests that the success of the result is probabilistic, as the application of the unitary operation on the state $|\psi\rangle$ will then be subject to a projection over the state $|0\rangle$ for a certain amount of ancilla qubits. Next, we describe the circuit for implementing the BE.

In order to exploit the sparse structure of $M$ we employ two operators: the 
column oracle $\hat{O}_c[M]$, that locates the $l$-th non null element in the $j$-th row in the column $c(j,l)$

\begin{align}
    \hat{O}_c|l\rangle_m|j\rangle_n= |l\rangle_m|c(j,l)\rangle_n,  
\end{align}
where $m=\lceil\log_2 s\rceil$, and the value oracle $\hat{O}_v[M]$,
\begin{align}
    \hat{O}_v[M]|0\rangle|l\rangle_m|j\rangle_n= \big( M_{j,c(j,l)}|0\rangle+\sqrt{1- M_{j,c(j,l)}^2}|1\rangle\big)|l\rangle_m|j\rangle_n,
\end{align}
that rotates an ancilla qubit intialised in $|0\rangle$ by the angle $\arccos[M_{j,c(j,l)}]$.

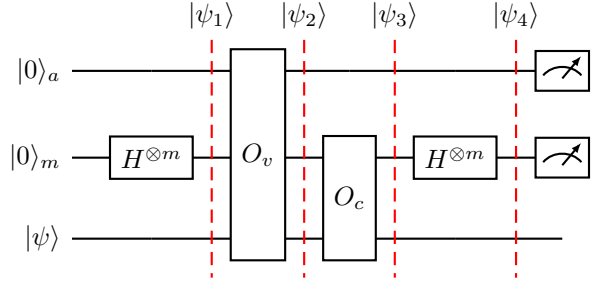
\begin{figure}[t]
  \centering
\begin{quantikz}
\lstick{$|0\rangle_a$}  &  \qw\slice{$|\psi_1\rangle$} &  \gate[3]{O_v} \slice{$|\psi_2\rangle$} &  \qw\slice{$|\psi_3\rangle$}&\qw\slice{$|\psi_4\rangle$}& \meter{} \\
\lstick{$|0\rangle_m$}  & \gate{H^{\otimes m}}&               & \gate[2]{O_c}&  \gate{H^{\otimes m}} & \meter{} \\
\lstick{$|\psi\rangle$} & \qw                 &               &	             &                   \qw & \qw
\end{quantikz}  
\caption{The circuit for the implementation of the linear operator $M$ using the block-encoding technique.}
  \label{fig:BE_circuit}
\end{figure}

From Ref.~\cite{camps_explicit_2024} we know that the operator $U_M$ can be implemented by applying the circuit of Fig.
~\ref{fig:BE_circuit},
which reads

\begin{align}\label{eq:unitary_be}
U_M=(\mathds{1}_2\otimes H^{\otimes m}\otimes \mathds{1}_N)(\mathds{1}_2\otimes\hat{O}_c)\hat{O}_M(\mathds{1}_2\otimes H^{\otimes m}\otimes \mathds{1}_N).
\end{align}

The Hadamard operations (first and last in Eq.~\eqref{eq:unitary_be}) transform $m$ 
ancilla qubits initialized in $|0\rangle$ into a superposition and viceversa.

The oracles, act as tools to implement the standard row-column matrix product, by exploiting the sparsity of the matrix.

Starting from the state $|\psi\rangle=\sum_i c_i|i\rangle$ we get by applying the operators one by one, cf.~Fig.\ref{fig:BE_circuit}:

\begin{align}
    |\psi_1\rangle &= \frac{1}{2^{\frac{m}{2}}}\sum_{l=0}^{2^m-1}\sum_{i=0}^{2^n-1}c_i|0\rangle|l\rangle|i\rangle\\
    |\psi_2\rangle &= \frac{1}{2^{\frac{m}{2}}}\sum_{l=0}^{2^m-1}\sum_{i=0}^{2^n-1}c_i\big(M_{i,c(i,l)}|0\rangle+\sqrt{1-M_{i,c(i,l)}^2}|1\rangle\big)|l\rangle|i\rangle\\
    |\psi_3\rangle &= \frac{1}{2^{\frac{m}{2}}}\sum_{l=0}^{2^m-1}\sum_{i=0}^{2^n-1}c_i\big(M_{i,c(i,l)}|0\rangle+\sqrt{1-M_{i,c(i,l)}^2}|1\rangle\big)|l\rangle|c(i,l)\rangle\\
    |\psi_4\rangle &= \frac{1}{2^m}\sum_{k,l=0}^{2^m-1}\sum_{i=0}^{2^n-1}(-1)^{kl}c_i\big(M_{i,c(i,l)}|0\rangle+\sqrt{1-M_{i,c(i,l)}^2}|1\rangle\big)|k\rangle|c(i,l)\rangle
\end{align}

Finally, the state $ M|\psi\rangle$ is recovered projecting the $m+1$ ancilla qubits on the state $|0\rangle_{m+1}$, yielding

\begin{align}\label{eq:explicit_M}
M|\psi\rangle=\sum_{l,i}c_iM_{i,c(i,l)}|c(i,l)\rangle.
\end{align}

There are a few caveats that we need to mention if we want to apply BE to recover the system sketched in Fig.~\ref{fig:carl_circuit}. First, all the Carleman operators $A,B, I\otimes A+A\otimes I,\dots$ have to be rescaled accordingly, so that the final state has norm 1.
Second, for each of them, we need to calculate the success probability and the number of ancilla qubits that are needed, and find those values such that the algorithm is optimized. 

In the following we are going to see how the BE works for the matrices $A$ and $B$.

\subsection{BE of matrix $A$}

We use the Euler forward method to evolve the system. The linear evolution of Eq.\eqref{eq:ADR_fe_costant_c} is encoded in the tridiagonal Toeplitz-Hankel matrix

\begin{align}\label{eq:linear_euler}
   L = \mathbb{1}+\Delta t A = \begin{pmatrix}
       \lambda_0 & \lambda_1             & 0                                      & \dots & \lambda_2\\
       \lambda_2             & \lambda_0 & \lambda_1& \ddots & 0 \\
        0                                                   & \ddots                                     & \ddots                          & \ddots & \vdots \\
        \vdots                                        & \ddots                                     &                                        &                  & \\
        \lambda_1              & 0                                                  & \dots                           & \lambda_2 & \lambda_0
    \end{pmatrix},
\end{align}
where we have introduced the variables 
\begin{align}
    \lambda_0&=1-2\gamma_{d}-\gamma_r,\nonumber\\
    \lambda_1&=\gamma_{d}-\frac{\gamma_a}{2},\nonumber\\
    \lambda_2&=\gamma_{d}+\frac{\gamma_a}{2},
\end{align}
with the strength of the diffusion, the advection and the linear reaction is given by the Courant numbers 
\begin{align}
    \gamma_d  = \frac{\Delta t D}{\Delta x^2} ,\quad\gamma_a =\frac{\Delta t U}{\Delta x},\quad \gamma_r=a \Delta t
\end{align}
respectively. Note that the ratios of the coefficients give the Damkohler and Peclet cell numbers as
\begin{align}
    \text{Da}_A=\frac{\gamma_a}{\gamma_r},\quad \text{Da}_D = \frac{\gamma_d}{\gamma_r},\quad \text{Pe}=\frac{\gamma_d}{\gamma_a}.
\end{align}
In order to apply the BE, $|\lambda_i|\leq 1$, for $i=0,1,2$ and the maximum eigenvalue $|\sum_i\lambda_i|\leq1$.
This results in conditions over the $\gamma$s:
\begin{align}
    |1-2\gamma_d-\gamma_r|&<1,\nonumber\\
   |\gamma_d-\gamma_a/2|&<1,\nonumber\\
   |\gamma_d+\gamma_a/2|&<1,\nonumber\\
   |1-\gamma_r|&<1,
\end{align}
which are usually satisfied, as each parameter $\gamma$ has to be much lower than 1 to ensure stability to the Euler's method.

\begin{figure}[t]
  \centering
\begin{quantikz}
\lstick{$|c\rangle_1$} &  \qw&\gate[3]{\hat{O}_c[A]}&\qw\\
\lstick{$|c\rangle_2$} &  \qw&&\qw\\
\lstick{$|\psi\rangle$} &  \qw&&\qw
\end{quantikz}
$=$
\begin{quantikz}
\qw&          &\ctrl{2}& \qw\\
\qw& \ctrl{1} & \qw&\qw\\
\qw& \gate{S_+}&\gate{S_-} &\qw
\end{quantikz}  \\
 \quad $S_+:$
\begin{quantikz}
\qw& \ctrl{2} &\ctrl{1}& \targ{}\\
\qw& \ctrl{1} &\targ{}&\qw\\
\qw& \targ{} &\qw &\qw
\end{quantikz}  
\quad $S_-:$
\begin{quantikz}
  \qw& \octrl{2} &\octrl{1}& \targ{}\\
 \qw& \octrl{1} &\targ{}&\qw\\
 \qw& \targ{} &\qw &\qw
\end{quantikz}  
\caption{The circuit for the implementation of the column oracle operator $\hat{O}_c[A]$ acting on the column register $|c\rangle$ composed by $m=2$ qubits. The operators $S_+$ and $S_-$ are the rightward and leftward shift operators, shown here for the case of the state $|\psi\rangle$ embedded on 3 qubits.}
  \label{fig:O_C_A_gates}
\end{figure}
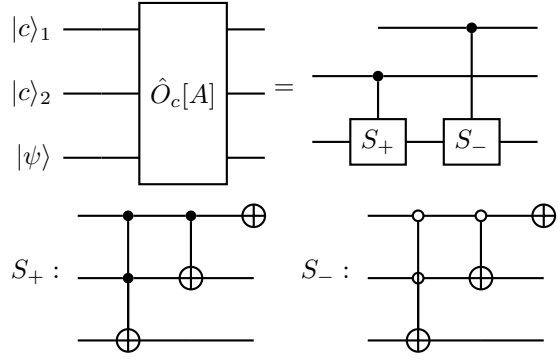

Matrix $A$ can be embedded into a unitary operation by using a total of 3 $(m=2)$ ancilla qubits. 
The sparsity of $A$ is 3, so we need to use 2 qubits to manage the column oracle. Following ref.~\cite{camps_explicit_2024} we define

\begin{align}
    c(i,0)&=c(i,3)=i,\nonumber\\
    c(i,1)&=\text{mod}[i+1,N],\nonumber\\
    c(i,2)&=\text{mod}[i-1,N],
\end{align}

\noindent where $\text{mod}[\cdot,N]$ is the modulus $N$ function. The corresponding oracle $\hat{O}_c$ can be implemented by two controlled operations $\text{C}S_{+}$ and $\text{C}S_{-}$, where $S_{+}$ and $S_{-}$ are the rightward and leftward shift operators~\cite{sanavio_lattice_2024}. The explicit gate expression is reported in Fig.~\ref{fig:O_C_A_gates}.

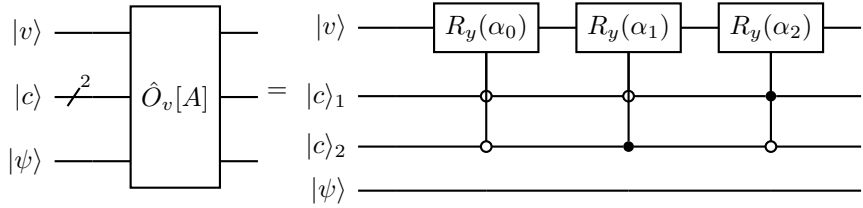
\begin{figure}[t]
  \centering
\begin{quantikz}
 \lstick{$|v\rangle$} & \qw&\gate[3]{\hat{O}_v[A]}&\qw\\
  \lstick{$|c\rangle$}&\qwbundle{2} &&\qw\\
 \lstick{$|\psi\rangle$} & \qw&&\qw
\end{quantikz}
$=$
\begin{quantikz}
 \lstick{$|v\rangle$} & \qw& \gate{R_y(\alpha_0)}&\gate{R_y(\alpha_1)}&\gate{R_y(\alpha_2)}&\qw\\
 \lstick{$|c\rangle_1$} & \qw& \octrl{-1}          & \octrl{-1}         &\ctrl{-1}&\qw\\
 \lstick{$|c\rangle_2$} & \qw& \octrl{-2}          & \ctrl{-2}          &\octrl{-2}&\qw\\
 \lstick{$|\psi\rangle$} & \qw& \qw                 &\qw                 &\qw       &\qw
\end{quantikz}  
\caption{The circuit for the implementation of the matrix oracle $\hat O_v[A]$. The rotations are controlled by the column register and applied on the value register. The last register encoding the state is not touched by the oracle, making it extremely efficient.}
  \label{fig:O_M_A_gates}
\end{figure}

The operator $\hat{O}_v[A]$ is obtained by applying a rotation of an angle $\alpha_l$ for each of the possible values of $l=0,1,2$. It is therefore described by the circuit in Fig.~\ref{fig:O_M_A_gates}

The values of the angles $\alpha_l$ can be obtained by solving a linear system as specified in \cite{camps_explicit_2024}. Moreover the circuit can be further simplified following Ref.~\cite{mottonen_quantum_2004} and approximated following Ref.~\cite{sanavio_imputation_2024} to write it as a combination of CNOT operators and single qubit rotations. 

The circuit proposed in this section is a combination of efficient circuits~\cite{camps_explicit_2024}, which makes use of the BE strategy by implementing efficient oracles that encode the information about the sparsity of the operators. 
However, this decomposition comes at a cost, which is the probabilistic nature of the result. In fact, we calculate the success probability $p_0[\hat M]$ for measuring $|0\rangle$ in the ancilla qubits after applying the operation $\hat M$  depending on the values of the relevant parameters $\gamma_{adv},\gamma_{diff},\gamma_{re}, N.$ Furthermore, $p_0$ depends also on the initial state of the system, as it is defined by

\begin{align}
    p_0[\hat M]=\frac{1}{2^{2m}}\lVert\sum_{i,l}c_iM_{i,c(i,l)}|c_i,l\rangle\rVert^2.
\end{align}

For the linear operation $L$ of Eq.\eqref{eq:linear_euler}, we get
    \begin{align}
        p_0[L]&=\frac{1}{16}\lVert\sum_{i=1}^Nc_i\big(\lambda_0|i\rangle+\lambda_1|i+1\rangle+\lambda_2|i-1\rangle\big)\rVert^2\\
        &=\frac{1}{16}\sum_i\bigg[|c_i|^2(\lambda_0^2+\lambda_1^2+\lambda_2^2)+(c_ic^{*}_{i+1}+c_ic^{*}_{i-1})(\lambda_0\lambda_1+\lambda_0\lambda_2)+(c_ic^{*}_{i+2}+c_ic^{*}_{i-2})\lambda_1\lambda_2\bigg].\nonumber
    \end{align}

\begin{figure}
 \begin{subfigure}{.5\textwidth}   \subcaption{}
    \centering
    \includegraphics[width=0.7\linewidth]{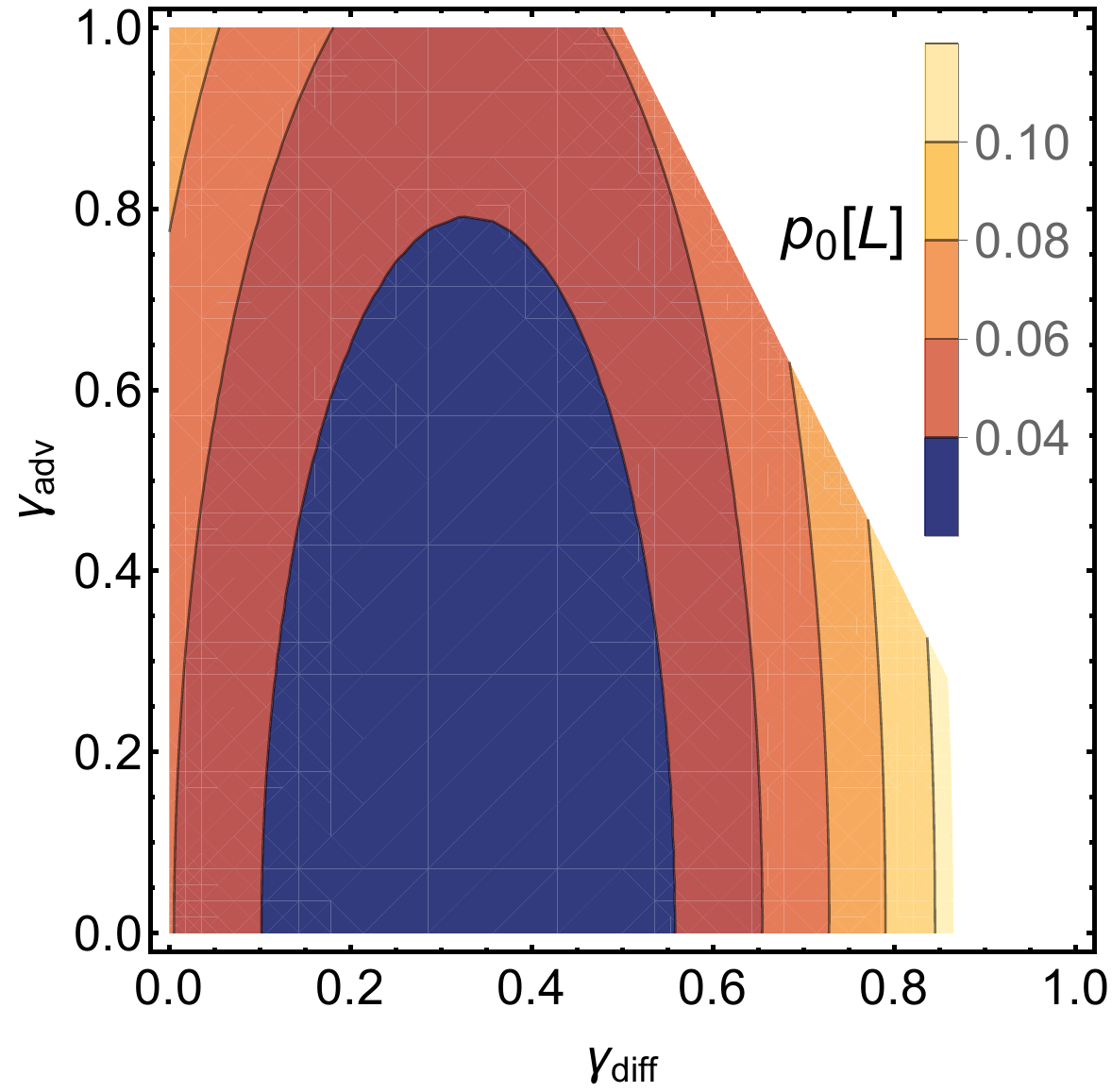}
  \end{subfigure}%
\begin{subfigure}{.5\textwidth}   \subcaption{}
  \centering
  \includegraphics[width=\linewidth]{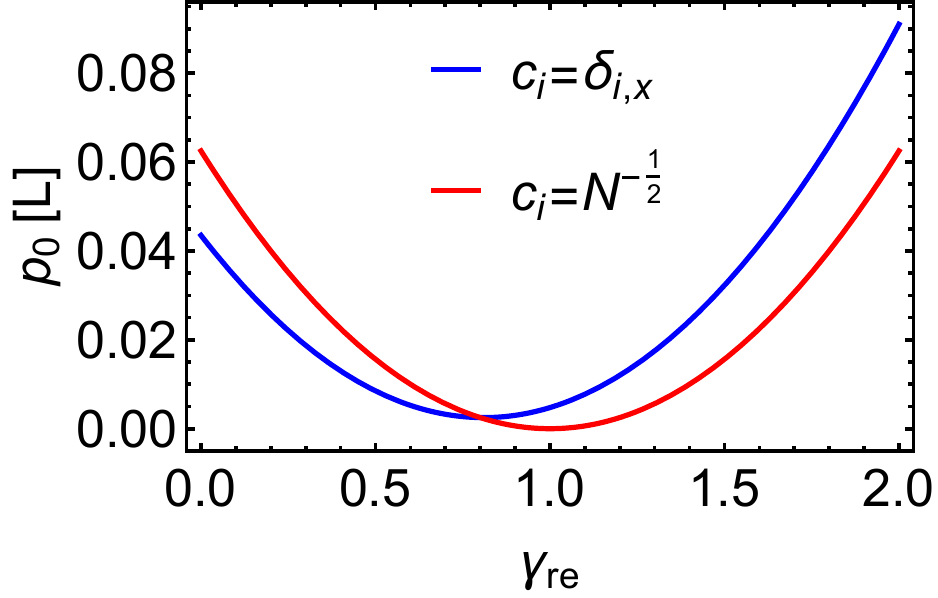}
\end{subfigure}
\caption{\label{fig:probability_a1}The success probability of implementing the linear operator $L$. In (a) plotted as a function of the coefficients $\gamma_{adv},\gamma_{diff}$ in the region where the the block-encoding implementation is possible. The parameters are set to $N=100, \gamma_{re}=0.01$. In (b) plotted as a function of $\gamma_{re}$ for the localized (blue line) and uniform (red line) cases. The parameters are set to $N=100, \gamma_{adv}=\gamma_{diff}=0.1$.}
\end{figure}

This expression is dependent on the initial state $|\psi\rangle.$ We plot in Fig.~\ref{fig:probability_a1}(a) the value of $p_0[L]$ as a function of $\gamma_{diff},\gamma_{adv}$ for an initial conditions where the state is localized on a single site $x$, with $c_i=\delta_{i,x}$, with the parameters set to $\gamma_{re}=0.01,$ $N=100$. In the other extreme case where the state is in uniformly spread over the whole chain, with $c_i=1/\sqrt{N}$, the probability $p_0[L]$ does not depend on the values $\gamma_{diff},\gamma_{adv}$ as it simplifies to
\begin{align}
    p_0[L]_{\text uniform}= \frac{1}{16}(1-\gamma_{re}^2)
\end{align}
In Fig.~\ref{fig:probability_a1}(b) we plot $p_0[L]$ as a function of $\gamma_{re}$ both in the localized case (blue line) with $\gamma_{adv}=\gamma_{diff}=0.1$ and in the uniform case (red line). 

From the results plotted in Fig.~\ref{fig:probability_a1}, we note that the success probability 
for the implementation of $L$ is low for realistic simulations, where $\Delta t$ is 
usually small, reaching at most the value 
$p_0[L]\approx0.12$ for $\gamma_{adv}=0.9, \gamma_{diff}=0.01$. 

Hence, the present circuit is able to reproduce {\it exactly} (ignoring the error of the single quantum gates) 
the linear ADR dynamics, but suffers of a poor success probability. 
Even setting $\Delta t =0$, the probability drops to $1/16$ due to the need of two ancilla qubits. 
This is in contrast with a unitary evolution, where the limit  $\Delta t \to 0$ 
smoothly recovers the identity operator, which one expects to implement with probability one. 

It is worth mentioning that in Ref.~\cite{huango_quantum_2024} the authors 
propose a circuit for the Advection-Diffusion equation using an unitary evolution in imaginary time. 
Their approach indeed converges to probability one in the limit $\Delta t\to 0$. 
However, the unitary evolution is implemented via a limited number of Trotter-steps, which reduces 
the overall accuracy of the dynamics. 
The authors report an upper bound of the success probability scaling as $e^{-1/\epsilon}$, 
$\epsilon$ being the final error, which represents a very severe limitation for 
any reasonable value of $\epsilon$.

Finally, it is also worth observing that in  Ref.~\cite{mezzacapo_quantum_2015},  a 
pioneering work for the implementation of an AD quantum circuit, the authors
report a single-step success probability well above $0.9$, thus shedding hopes 
for the practical application of the circuit. 
However, a comparison between their work and the present one is difficult because 
the two methods are based on pretty different techniques. 
For one, they considered a Lattice Boltzmann~\cite{benzi_lattice_1992} implementation of the AD equation, 
based on a second-quantized representation of the LB  streaming-relaxation 
dynamics~\cite{benzi_lattice_1992}, along with a linear combination of unitaries~\cite{childs_quantum_2017}.
This might indeed result in higher success probability of the non-unitary update, as also observed 
for Lattice Boltzmann based Carleman procedure for simulation of fluids~\cite{itani_analysis_2022,itani_quantum_2024,sanavio_quantum_2024}.
However, based on the current experience for the case of fluids, such
an approach is most likely to face an exponential depth of the quantum circuit
with the number of qubits ~\cite{sanavio_lattice_2024}.
Since this aspect is not reported in Ref.~\cite{mezzacapo_quantum_2015}, the viability
of the corresponding quantum algorithm cannot be assessed.

\subsection{BE of matrix $B$}

The circuit for implementing $B$ has to be applied after the circuit for $A$.  
Because of the linearity of quantum mechanics, the circuit has to implement the linear operation 

\begin{align}
    \hat{B} &=\begin{pmatrix}
        I_n & B\\
        0 & I_{2n}
    \end{pmatrix},
\end{align}
where $I_n$ is the identity on the space of $n$ qubits.
Because the matrix $\hat B$ is 2-sparse, we need only 1 qubit to embed the information about the position of the column entry. The column operator  $\hat{O}_c[B]$ is such that

\begin{align}
    c_B(i,0)&=i,\\
    c_B(i,1)&=\text{mod}[i+(i+1)N,N+N^2].\nonumber
\end{align}

In order to define its circuital expression, we see that for each $|i\rangle$, the oracle $\hat O_c[B]$ applies a right shift to the state $|i+(i+1)N\rangle$, hence it can be obtained as a combination of two circuits. The first circuit changes the order of the Carleman states of second order, by transforming 
\begin{align}
    |i+(i+1)N\rangle\to|N+i\rangle, \text{ for } i>N.
\end{align}
This can either be obtained with a combination of swap gates that sorts the local quadratic components, or the state can be initialized in this order at the beginning. Finally, a controlled rightward shift $CS_+^{N}$ applied $N$ times  brings the state in the wanted position. In Ref.~\cite{camps_explicit_2024} it is shown that any power of the shift operator can be efficiently implemented.  

The matrix operator $\hat{O}_v[B]$ has the form

\begin{align}
    \hat{O}_v[B]|0\rangle|0\rangle|i\rangle&=|0\rangle|0\rangle|i\rangle\\
    \hat{O}_v[B]|0\rangle|1\rangle|i\rangle&=\begin{cases}
        R_y[\beta]|0\rangle|1\rangle|i\rangle&\quad\text{if }i<N\\
        |1\rangle|1\rangle|i\rangle&\quad\text{if }i\geq N
    \end{cases}\nonumber
\end{align}

\noindent where $\beta =2\arccos(b\Delta t).$  The implementation of $\hat O_v[B]$ follows the same strategy as the one of $\hat O_v[A]$, but it requires an extra quantum register to implement the comparison circuit~\cite{donaire_lowering_2024}. A rotation by an angle $\beta$ or $\pi/2$ is applied if the state $|i\rangle$ has label either lower or greater than $N$. 

The application of Eq.~\eqref{eq:explicit_M} leads to a success probability of the circuit for the quadratic interaction at most of
\begin{align}
    p_0[B]=\frac{1}{4}(1+\Delta t^2 b^2),
\end{align}
\noindent with realistic values of $\Delta t$ and $b$ yielding to $p_0[B]\approx 0.25$. 
This reduces the probability of a successful implementation of the algorithm by a factor 4 w.r.t. the probability obtained for the application of $L$, being the two independent outcomes. 

\section{Conclusion and Outlooks}
We analysed the feasibility of a quantum algorithm for the one-dimensional advection-diffusion-reaction (ADR) 
equation with a logistic reaction term. This describes an evolution problem that is both nonlinear and non-unitary. 
Our analysis shows that the Carleman embedding provides an efficient strategy to linearise the 
discretised ADR equation, since it shows an exponential convergence to the solution 
of the nonlinear system even in the presence of numerical fluctuations at high Peclet numbers. 
Nonetheless, finding a strategy to efficiently solve the linearised problem on a quantum computer is still an open challenge.
We showed that a straight decomposition of the non-unitary Carleman matrix into the Pauli 
demands a number of quantum gates that scales exponentially with the number of sites. 
By exploiting matrix sparsity and using efficient explicit matrix-access 
oracles, block-encoding we have been able to reduce the gate complexity 
of the circuit from exponential to polynomial. 
However,  the probabilistic nature of the technique poses a significant challenge because the success
probability of the single-step formulation is significantly below the unit value.
Further research is needed on the encoding of the linearised non-unitary matrix into a quantum circuit that 
would enable the practical implementation of Carleman quantum algorithms for advection-diffusion
reaction problems.  

\bmhead{Acknowledgements}

We acknowledge financial support from National Centre for HPC, Big Data and Quantum Computing (Spoke 10, CN00000013).
The authors gratefully acknowledge discussions with A. Ralli and P. Love.


\end{document}